\documentclass[manuscript]{acmart}
\AtBeginDocument{%
  }

\setcopyright{none}                                

\usepackage{tikz}
\usetikzlibrary{shapes,backgrounds}
\usepackage{amsmath}
\usepackage{hyperref}
\usepackage{url}
\usepackage{threeparttable} 

\usepackage{filecontents}
\usepackage{tabularray}
\UseTblrLibrary{booktabs}

\usepackage{subcaption} 
\usepackage{graphicx} 

\newcommand{\URoman}[1]{\uppercase\expandafter{\romannumeral#1}}

\begin{document}

\title{Accelerating Recommender Model ETL with a Streaming FPGA–GPU Dataflow}








\author{Yu Zhu}
\affiliation{%
  \institution{ETH Zurich}
  \city{Zurich}
  \country{Switzerland}}
\email{yu.zhu@inf.ethz.ch}

\author{Wenqi Jiang}
\affiliation{%
  \institution{ETH Zurich}
  \city{Zurich}
  \country{Switzerland}}
\email{wenqi.jiang@inf.ethz.ch}

\author{Piyumi Jasin Pathiranage}
\affiliation{%
  \institution{ETH Zurich}
  \city{Zurich}
  \country{Switzerland}}
\email{ppathiran@ethz.ch}

\author{Yongjun He}
\affiliation{%
  \institution{ETH Zurich}
  \city{Zurich}
  \country{Switzerland}}
\email{yongjun.he@inf.ethz.ch}

\author{Gustavo Alonso}
\affiliation{%
  \institution{ETH Zurich}
  \city{Zurich}
  \country{Switzerland}}
\email{alonso@inf.ethz.ch}

\renewcommand{\shortauthors}{Yu Zhu et al.}

\begin{abstract}
The real-time performance of recommender models depends on the continuous integration of massive volumes of new user interaction data into training pipelines.  
While GPUs have scaled model training throughput, the data preprocessing stage---commonly expressed as Extract--Transform--Load (ETL) pipelines---has emerged as the dominant bottleneck.  
Production systems often dedicate clusters of CPU servers to support a single GPU node, leading to high operational cost.  
To address this issue, we present \textsc{PipeRec}, a hardware--accelerated ETL engine co-designed with online recommender model training.  
\textsc{PipeRec} introduces a training-aware ETL abstraction that exposes freshness, ordering, and batching semantics while compiling software-defined operators into reconfigurable FPGA dataflows and overlaps ETL with GPU training to maximize utilization under I/O constraints.  
To eliminate CPU bottlenecks, \textsc{PipeRec} implements a format-aware packer that streams training-ready batches directly into GPU memory via P2P DMA transfers, enabling zero-copy ingest and efficient GPU consumption.  
Our evaluation on three datasets shows that \textsc{PipeRec} accelerates ETL throughput by over 10× compared to CPU-based pipelines and up to 17$\times$ over state-of-the-art GPU ETL systems.  
When integrated with training, \textsc{PipeRec} maintains 64--91\% GPU utilization and reduces end-to-end training time to 9.94\% of the time taken by CPU--GPU pipelines.
\end{abstract}




\maketitle

\section{Introduction}

Data preprocessing is a critical component of modern machine learning (ML) pipelines.  
Raw data from diverse sources and formats must be transformed through Extract--Transform--Load (ETL) pipelines---including normalization, feature encoding, augmentation and other techniques---before it can be used for model training~\cite{jain2020overview, chen2021data, he2021automl, li2021cleanml, budach2022effects, qi2023auto, li2023volcanoml}.  
In today’s ML systems, CPUs typically execute ETL, after which the processed data is transferred to GPUs for training~\cite{fan2004gpu, zhang2020understanding, choquette2021nvidia, mudigere2022software, choquette2023nvidia, tirumala2024nvidia}.  
For models that require frequent updates, such as recommender systems, ETL must be executed online~\cite{ramirez2017survey, egg2021online, sima2022ekko, liu2022monolith, qin2024datasetgrowth} to meet strict service-level objectives (SLOs).  

\begin{figure}[t]
    \centering 
    \begin{subfigure}[b]{0.55\textwidth} 
        \includegraphics[width=1\linewidth]{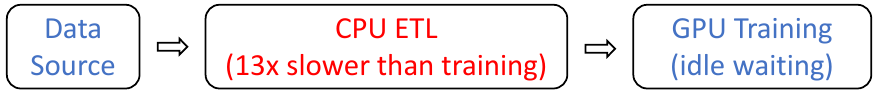}
        \caption{End-to-end pipeline schematic: CPU ETL dominates the runtime, leaving the GPU idle.}
        \vspace{0.5em}
        \label{fig:preprocess_schematic}
    \end{subfigure}
    \begin{subfigure}[b]{0.6\textwidth} 
        \includegraphics[width=1\linewidth]{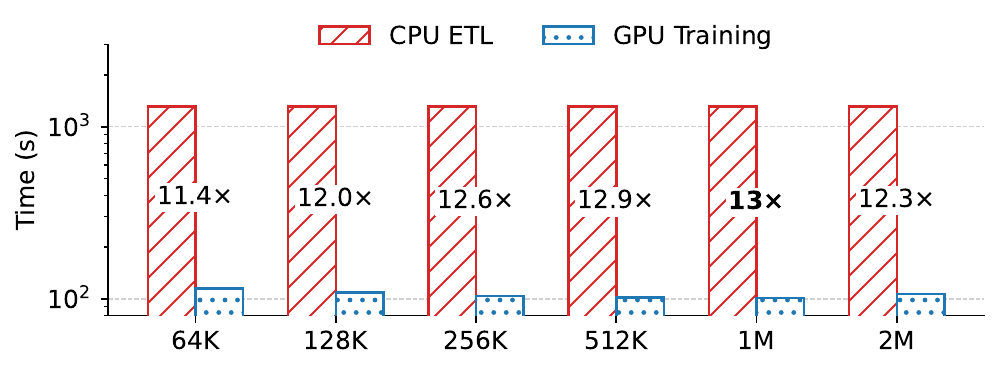}
        \caption{Stage runtimes: CPU ETL for data preprocessing is up to 13× slower than training across batch sizes.}
        \label{fig:preprocess_train_min}
    \end{subfigure}
    \begin{subfigure}[b]{0.6\textwidth} 
        \includegraphics[width=1\linewidth]{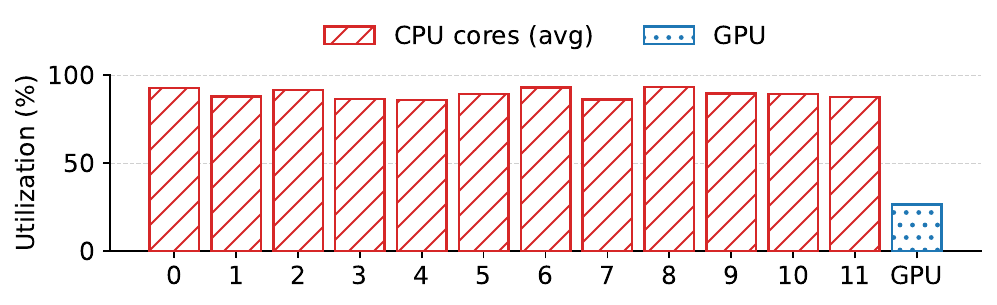}
        \caption{Resource utilization: all 12 CPU cores are saturated, while Nvidia A100 remains underutilized.} 
        \label{fig:preprocess_cpu_gpu}
    \end{subfigure}
    \caption{
        The ETL bottleneck in a CPU-based commercial DLRM pipeline, leaving the GPU underutilized.
    }
    \label{fig:preprocess_train}
\end{figure}

\begin{figure*}[t]
    \centering
    \includegraphics[width=0.98\linewidth]{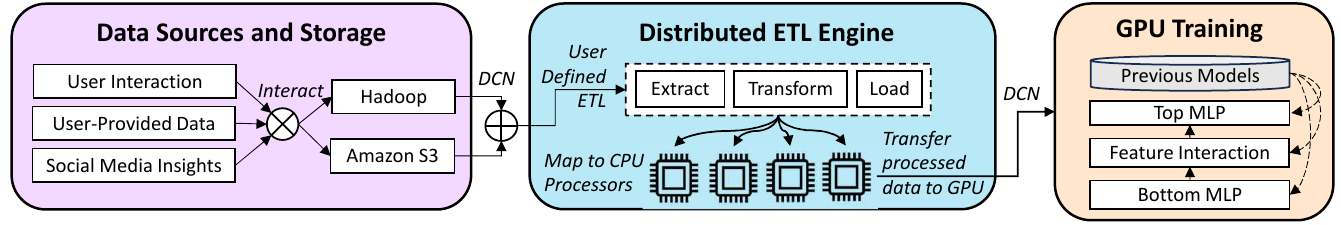}
    \caption{Typical architecture of continuous training for recommender models in CPU–GPU clusters. Raw data from storage is ingested via the data center network (DCN) into a distributed ETL engine, where user-defined transformations execute on CPUs. Processed batches are then transferred to GPUs for model training, often creating a bottleneck between ETL and training.}
    \label{fig:overall_dataflow}
\end{figure*}

As GPUs have scaled dramatically in compute capacity~\cite{dally2021evolution}, CPU-based ETL has emerged as the main bottleneck in end-to-end ML pipelines \cite{graur2022cachew, graur2024pecan, Bother2025Modyn, zhao2023goldminer, kuchnik2022plumber, murray2021tf, audibert2023tf}.  
This imbalance is especially severe for recommender models \cite{zhao2021understanding}, which are relatively small~\cite{gupta2020architectural, jiang2021fleetrec}, leaving GPUs underutilized while CPUs struggle with preprocessing demands.  
Figure~\ref{fig:preprocess_train} summarizes a production Deep Learning Recommender Model (DLRM) pipeline \cite{zhao2022understanding} and quantifies the imbalance between CPU-side ETL and GPU training  in Nvidia A100 with 12 CPU cores.
Figure~\ref{fig:preprocess_schematic} shows the pipeline stages, with CPU ETL sitting on the critical path between the data source and the GPU.
Figure~\ref{fig:preprocess_train_min} reports per-epoch stage times across batch sizes (64K--2M) on a logarithmic scale, where CPU ETL is consistently 11.4$\times$--13.0$\times$ slower than training and thus contributes over 90\% of the end-to-end wall-clock time.
Figure~\ref{fig:preprocess_cpu_gpu} shows average resource utilization: all 12 CPU cores are saturated, while the GPU remains largely idle (roughly 10--15\% utilization), waiting for data to arrive.
Taken together, these measurements indicate that CPU-bound ETL rather than GPU compute dominates time-to-train, and that simply provisioning faster GPUs provides little benefit without addressing the ETL stage.
This observation motivates our design goal: \textit{to rebalance the pipeline by eliminating the ETL bottleneck and converting idle GPU cycles into productive training.}

Existing distributed CPU-based preprocessing services~\cite{murray2021tf, audibert2023tf, graur2022cachew, zhao2022understanding, zhao2023goldminer} scale throughput by adding servers, but consume significant resources and can account for over 60\% of total training energy~\cite{meta_dpp, zhao2022understanding}.  
GPU-based preprocessing systems such as NVTabular~\cite{nvidia_nvtabular_dlrm} achieve higher throughput but compete with training for GPU resources and are still power-inefficient.  
This motivates exploring dedicated hardware accelerators like FPGA as the ETL engine, where reconfigurable pipelines can deliver high throughput at much lower energy cost without interfering with GPU training.

We propose \textsc{PipeRec}, an FPGA--GPU co-designed solution that tightly integrates
a hardware-accelerated ETL engine with continuous training. 
At the programming interface level, \textsc{PipeRec} provides a compilation flow 
that maps software-defined ETL operators to hardware modules, applies operator fusion, 
and allocates state across memory hierarchies. 
At the system level, user-defined preprocessing pipelines are compiled into parallel 
streaming dataflows on FPGAs, which ingest data directly from memory, storage, or the network 
and transform it at line rate. 
At the runtime level, a lightweight CPU control plane orchestrates execution, while the FPGA 
performs both ETL and peer-to-peer transfers into GPU memory, eliminating host-side bottlenecks. 
At the I/O and deployment level, an integrated memory subsystem manages data movement 
across on-board, host, and remote memory and applies backpressure to balance the throughput 
between ETL and training, sustaining high GPU utilization. 
For scalability, stateless operators scale through replication, stateful operators share tables 
across lanes, and multiple pipelines can be instantiated concurrently on a single accelerator. 
This design allows ETL and training to proceed in parallel, with the GPU training on one batch while the FPGA prepares the next one.  
We prototype \textsc{PipeRec} on FPGA hardware and evaluate it using ETL pipelines from commercial DLRM workloads \cite{meta_dlrm, tf_dlrm}.  
Across three representative datasets, \textsc{PipeRec} consistently outperforms software-based preprocessing: it delivers one to three orders of magnitude higher throughput than multi-core CPU pipelines and achieves 2.4-17$\times$ speedups over GPU-based ETL, while lowering power consumption by a factor of 2.9$\times$--6.4$\times$.  
More importantly, by streaming batches directly into GPU memory and overlapping ETL with training, \textsc{PipeRec} sustains 64--91\% GPU utilization, in contrast to CPU-based designs that leave accelerators frequently idle.  
This leads to an overall reduction in training time of 10.06$\times$, demonstrating that ETL acceleration can shift the end-to-end performance bottleneck in recommender training.

\textbf{Contributions.}  
The paper makes the following contributions:
\begin{itemize}
  \item A training-aware ETL abstraction that bridges software-defined preprocessing operators and FPGA pipelines while providing batching semantics.  
  \item A co-scheduling runtime that pipelines FPGA-based ETL with GPU training to maximize GPU utilization under I/O constraints.  
  \item A format-aware packer that streams training-ready batches directly into GPU memory via P2P DMA, enabling zero-copy ingest.  
  \item A prototype of PipeRec on FPGAs, evaluated on large-scale DLRM datasets, demonstrating its performance and energy efficiency compared to CPU and GPU solutions.  
\end{itemize}

\section{Background and Motivation}\label{sec:background}

Online recommendation systems train continuously on streams of user interactions, as model freshness is as relevant as raw throughput.
Unlike offline workloads operating on static snapshots, continuous training must ingest, transform, and learn from event logs in near real time, while ensuring that each training sample only uses information available at the exact time of the event (point-in-time correctness) and avoiding training--serving skew.
Modern production pipelines therefore couple streaming ETL with iterative training, warm-starting from previous checkpoints, and updating large sparse embeddings alongside relatively small MLP stacks.
This setting stresses heterogeneous compute: CPUs execute join/aggregate/encode operations over high-cardinality features, while GPUs execute forward/backward passes and embedding lookups at millisecond to second timescales.
As Figure~\ref{fig:overall_dataflow} illustrates, the end-to-end path spans data sources, ETL, and the GPU trainer, and any rate mismatch forces GPUs to idle or ETL queues to back up.
In practice, CPU-bound ETL dominates wall-clock time for DLRM-style models, saturating cores while leaving accelerators underutilized.
These characteristics motivate moving part of the ETL closer to an accelerator, exposing a scheduled DAG that can stream preprocessed batches at line rate. 
By co-designing ETL with the trainer and enabling direct PCIe/RDMA paths between ETL engines and GPUs, the system reduces data movement, eliminates CPU bottlenecks, and improves time-to-freshness for online models.

\subsection{Dynamic Datasets}
Learning datasets in recommender systems are often dynamic, as they continuously grow with the collection of new samples \cite{song2014online, freno2017practical, eirinaki2018recommender, zhou2021contrastive, guo2023evaluating} experiencing both data evolution and data drift \cite{egg2021online, Bother2025Modyn}. 
Data evolution refers to natural changes such as the introduction of new features, classes, or an increase in data volume; data drift involves changes in the distribution of inputs, labels, or the relationships between inputs and outputs, leading to model performance degradation. 
To address these challenges, it is crucial to detect such changes, retrain or update models as needed, and ensure that data pipelines remain robust to maintain the accuracy and reliability of models in dynamic environments. 
For instance, daily retraining of models at the GrubHub food delivery platform can lead to a 20\% increase in purchase rates compared to not retraining the models \cite{egg2021online}. 
Similarly, a high-performance data loader can greatly improve the end-to-end system performance \cite{zolnouri2020importance, cai2020fly, bai2021efficient, ofeidis2022overview, jia2022data, svogor2022profiling}, highlighting the need for efficient ETL pipelines.  

\begin{table}[t]
  \caption{Transformation operators for different features, extracted from publicly available repositories and papers \cite{tf_dlrm, meta_dlrm, zhao2021understanding}.} 
  \label{tab:operator_list}
  \small
  \begin{tabular}{lll}
    \toprule
    \textbf{Operators} & \textbf{Description} & \textbf{Category} \\
    \midrule
    OneHot & Apply one hot encoding to normalize & dense, stateless\\
    Clamp & Restrict values within a specified range. & dense, stateless  \\
    Logarithm & Do log(x+1) operation. & dense, stateless  \\ 
    Hex2Int & Convert hex strings to decimal. & sparse, stateless   \\
    Modulus & Compute positive modulus. & sparse, stateless  \\
    Cartesian & Compute Cartesian product & sparse, stateless\\
    SigridHash & Compute hash value to normalize list & sparse, stateless\\
    VocabGen & Create tables from unique values. & sparse, stateful  \\
    VocabMap & Map values to generated indices. & sparse, stateful  \\ 
    Bucketize & Shard features based on bucket borders & both, stateless\\
    FillMissing & Fill missing elements with default values & both, stateless \\
  \bottomrule
  \end{tabular} 
\end{table}

\subsection{ETL for Recommender Systems}

Recommender pipelines ingest heterogeneous signals---numerical fields, categorical tokens, text, and images---and convert them into model-ready tensors.
Training ultimately operates on \emph{embeddings}, i.e., low-dimensional continuous vectors that represent input features \cite{zhao2023embedding}.
The ETL stage therefore extracts raw fields, applies feature-specific transformations, and emits dense tensors and integer indices that drive embedding lookups during training.

Input features are commonly partitioned into \emph{dense} and \emph{sparse} groups \cite{gupta2020architectural, cheng2016wide, naumov2019deep}.
Dense features (e.g., user age or item price) are well-defined numerical attributes that are normalized to improve optimization stability.
Standard cleaning and scaling include clipping invalid or negative values, applying a logarithmic transform to reduce skew, and performing z-score normalization.
Sparse features originate from high-cardinality categorical data and may contain missing values.
Typical examples include user IDs, item IDs, and advertisement categories, which must be mapped to compact integer spaces before embedding.

Table~\ref{tab:operator_list} lists the operators used for DLRM in open-source pipelines published by Meta and Google \cite{meta_dlrm, tf_dlrm, zhao2021understanding}.
Each operator performs a relatively small transformation and the operators are chained to achieve the desired result: 
\emph{OneHot} encodes small-cardinality bins as indicators (e.g., \texttt{bin=3, K=5$\rightarrow$[0,0,0,1,0]});
\emph{Clamp} restricts values to a range (e.g., \texttt{x=-1,[0,10]$\rightarrow$0});
\emph{Logarithm} reduces skewness and compresses heavy tails (\texttt{x=999$\rightarrow\log(999+1)$});
\emph{Hex2Int} canonicalizes hex strings (e.g., \texttt{"0x1a3f"$\rightarrow$6719});
\emph{Modulus} maps IDs to a bounded range (e.g., \texttt{(-7) mod 5$\rightarrow$3}); 
\emph{Cartesian} forms cross features (e.g., \texttt{(user\_id=42, ad\_id=17) $\rightarrow$ "42|17"} or \texttt{hash(42,17) mod $M$}) to create a new categorical key distinct from the originals.
\emph{SigridHash} bounds categorical IDs (\texttt{hash(id)\%M});
\emph{VocabGen/Map} provide persistent token$\!\to$
index mapping for embedding lookups; 
\emph{Bucketize} discretizes a scalar by bin boundaries (e.g., \texttt{x=37, bins=[10,20,40]$\rightarrow$}\texttt{bin 3}),
and \emph{FillMissing} imputes NaNs (e.g., \texttt{[3.2, NaN]$\rightarrow$[3.2, 0.0]}).

In summary, the ETL pipeline bridges raw logs and the training loop by producing normalized dense tensors and indexed sparse features, thereby enabling efficient embedding lookups and stable optimization for training large-scale recommender models.

\begin{figure*}[!t]
    \centering
    \includegraphics[width=0.95\linewidth]{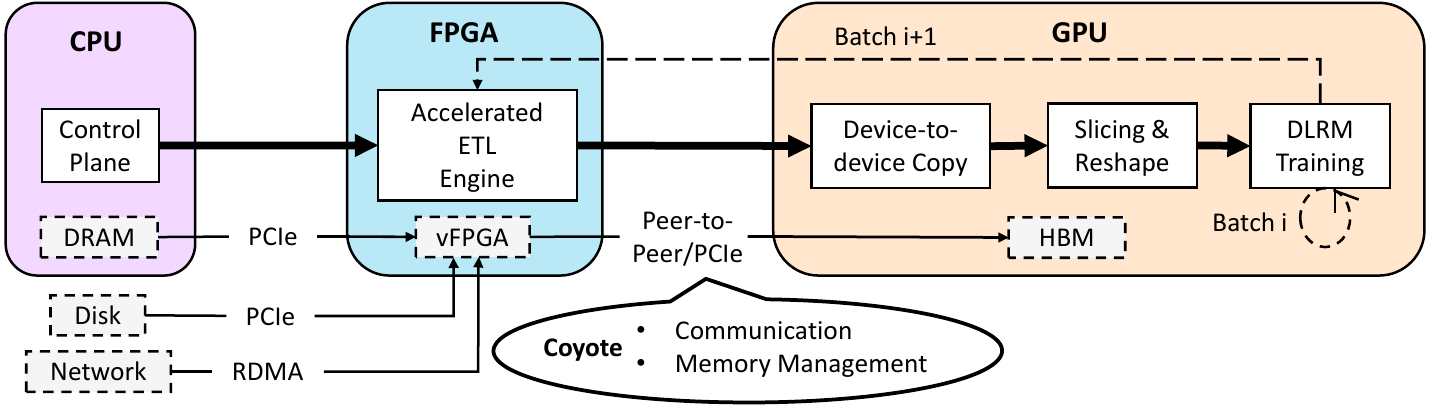}
    \caption{PipeReC: The CPU runs a light control plane; an FPGA implements the ETL engine and streams preprocessed batches via peer-to-peer PCIe to the GPU, where device-to-device copy, slicing/reshape, and DLRM training overlap in time (batch i training, batch i+1 ingest). 
    }
    \label{fig:fpga-gpu-codesign}
\end{figure*}

\subsection{ETL Platforms}

ETL platforms for recommender systems must process large, structured logs under strict freshness and cost constraints.
CPUs are often used in production due to their flexibility and mature ecosystems \cite{zhao2022understanding, zhao2023goldminer}.
Framework APIs such as TensorFlow \texttt{tf.data} and PyTorch \texttt{DataLoader}/\texttt{Dataset} support streaming inputs, parallel workers, and asynchronous pipelines, while domain libraries (e.g., \texttt{tf.image} and \texttt{torchvision}) provide optimized operators—CPU by default with GPU kernels available only for a subset of transforms (notably, decoding often remains CPU-bound).
Columnar formats (e.g., Parquet \cite{vohra2016apache}) over HDFS \cite{hadoop} and Amazon S3 \cite{amazon_s3} enable selective access to features and efficient scans.
Despite these advantages, CPU-based ETL often scales poorly for DLRM-style pipelines, requiring many servers to saturate a single GPU \cite{zhao2022understanding, zhao2023goldminer}.
In deployed systems, ETL remains predominantly CPU-bound, incurring extra data movement and large core counts across multiple machines to keep accelerators busy \cite{zhao2022understanding, zhao2023goldminer}.
Meanwhile, training throughput on modern GPUs continues to rise via reduced-precision arithmetic (BF16/FP16/FP8) and architectural advances \cite{markidis2018nvidia, nandakumar2018mixed, kalamkar2019study, sun2019hybrid, cherubin2020tools}, widening the rate gap between ETL and training and leading to idle accelerators and longer time-to-freshness for online models.

GPU-based ETL (e.g., DALI \cite{nvidia_dali}, Merlin/NVTabular \cite{oldridge2020merlin}, RAPIDS \cite{nvidia_rapids}) delivers high throughput by exploiting massive parallelism, high-bandwidth memory, and features such as GPUDirect Storage \cite{gpu_direct_storage}.
However, when ETL and training share the same accelerators, they contend for scarce HBM capacity and SM cycles, which can depress up to 10\% end-to-end throughput when compared against training-only baselines \cite{wang2024rap}.

A complementary architecture offloads ETL to reconfigurable logic (FPGAs), deployed as network- or PCIe-attached devices that stream preprocessed batches directly into GPU memory via DMA/RDMA.
This decouples preprocessing capacity from GPU training, removes contention for HBM/SM resources, and enables deeply pipelined dataflows with predictable latency and favorable performance-per-watt \cite{boutros2024field}.
Because ETL demand scales with data volume rather than model size, FPGA stages can be sharded and scaled independently, while GPUs are reserved for forward/backward passes and embedding lookups.
The principal trade-offs are the need to compile high-level ETL operators into hardware dataflows and to maintain equivalent functions with software stacks, a procedure presented in subsequent sections.

\section{\textsc{PipeRec:} Hardware Accelerated ETL Engine}\label{sec:fpga-gpu}

Figure~\ref{fig:fpga-gpu-codesign} presents the end-to-end architecture with a light CPU control plane, an FPGA ETL data plane, and a GPU training backend.
The CPU configures schemas, vocabularies, and buffer descriptors, but is not on the data path.
The FPGA implements a scheduled ETL dataflow that extracts, transforms, and packs feature tensors into training-ready batches.
Preprocessed batches are streamed directly into GPU HBM via P2P (peer-to-peer) PCIe, avoiding copies through host DRAM.
On the GPU, a device-to-device placement followed by a slicing/reshape step materializes the framework tensors and launches the DLRM forward/backward pass.
Double buffering overlaps stages so that batch $i$ trains while batch $i{+}1$ is ingested and reshaped, eliminating idle periods when the ETL side keeps pace.
Backpressure is explicit: the FPGA writes only when the GPU notifies a free staging buffer, which rate-matches ingestion to trainer consumption.
The open-source Coyote \cite{korolija2020abstractions} provides the communication and memory-management substrate across PCIe/RDMA, handling buffer registration, address exchange, and completions.
This separation of concerns decouples preprocessing capacity from the trainer, enables deterministic streaming dataflows, and minimizes CPU involvement on the steady-state path.

\begin{figure*}[t]
    \centering
    \includegraphics[width=0.85\linewidth]{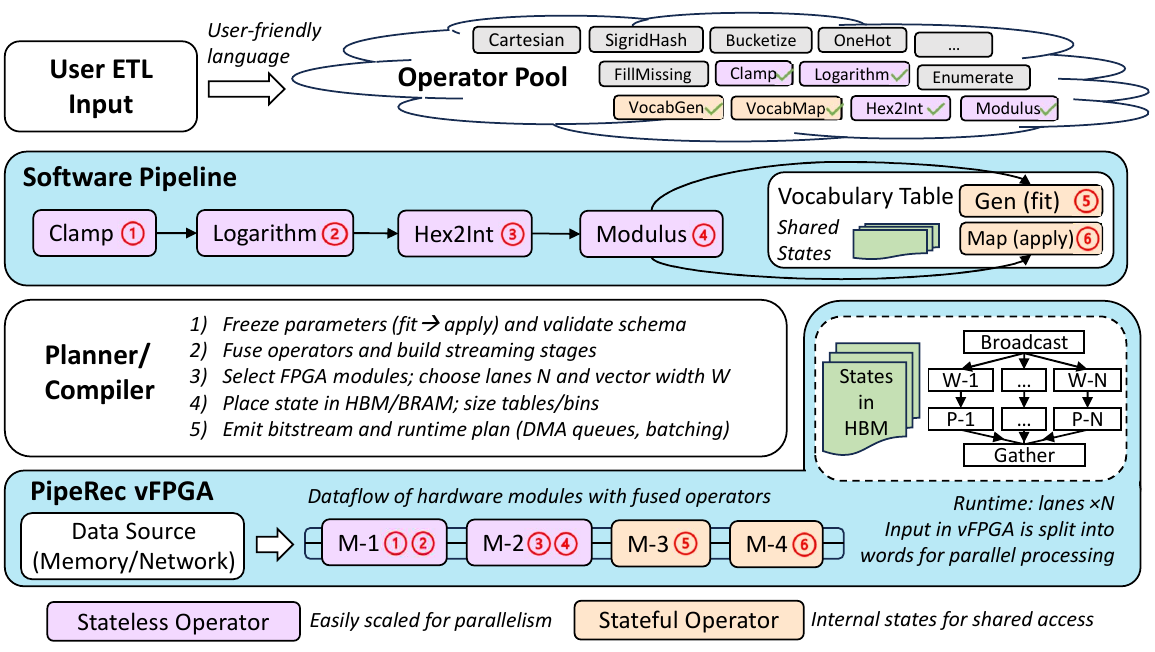}
    \caption{Compiling ETL pipelines from an operator pool to a streaming FPGA dataflow.}
    \label{fig:etl_vfpga}
\end{figure*}

\subsection{DAG and Mapping in FPGA}\label{sec:dataflow}

Figure~\ref{fig:etl_vfpga} summarizes how user ETL pipelines are compiled into a streaming FPGA dataflow.
Pipelines are expressed over an operator pool that includes both stateless transforms (e.g., \textit{Clamp}, \textit{Logarithm}, \textit{Hex2Int}, \textit{Modulus}) and stateful transforms (e.g., \textit{VocabGen}/\textsc{VocabMap}).
The software pipeline is validated against a schema and separated into a \emph{fit} phase (to learn operator parameters and vocabularies) and an \emph{apply} phase (to transform the stream using frozen parameters).
The planner–compiler then generates a hardware plan in five steps:
(1) freeze operator parameters and verify type/shape constraints;
(2) fuse compatible operators into streaming stages to minimize buffering and control overheads;
(3) select hardware modules for each stage and choose the degree of parallelism via the number of lanes $N$ and vector width $W$;
(4) place state in on-chip BRAM or HBM and size tables/bins for expected key distributions; and
(5) emit a bitstream together with a runtime plan that includes DMA queue layouts, batching policy, and buffer descriptors.
Stateless operators are replicated across lanes to scale throughput, whereas stateful operators expose shared state and are connected through broadcast/gather fabrics so that many lanes can access common tables without duplication.
At runtime the \textit{vFPGA} presents a dataflow of fused modules (\textit{M-1, M-2, \dots}) fed by a memory/network source; input words are split into $W$-wide vectors, processed in parallel across $N$ lanes, and re-packed into GPU-ready batches.
This organization preserves the logical operator semantics while delivering line-rate streaming with deterministic latency.

\begin{figure}[t]
    \centering
    \includegraphics[width=0.7\linewidth]{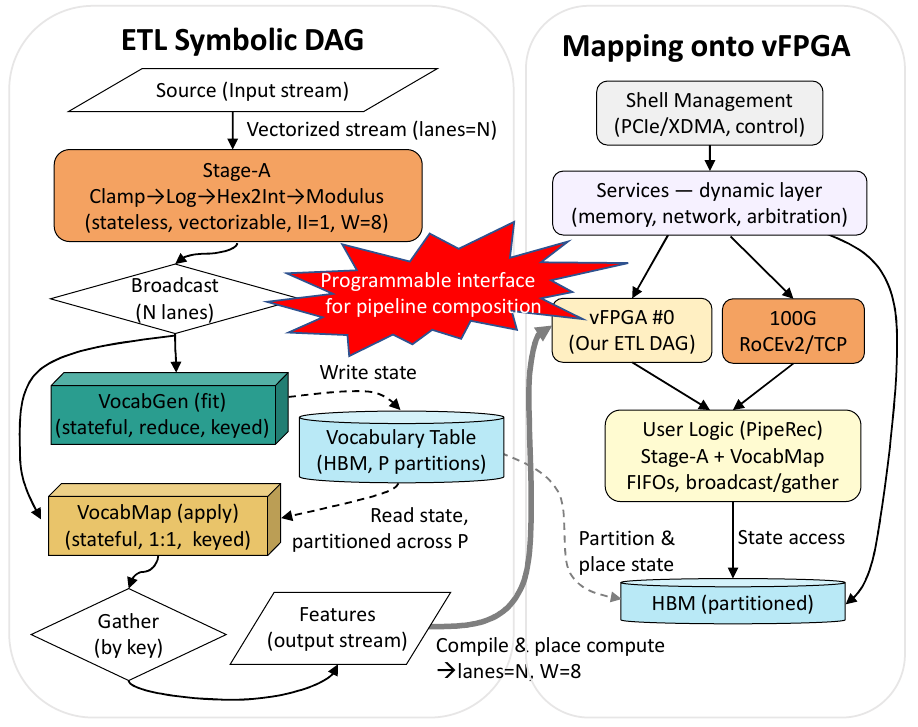}
    \caption{Automatically map from ETL symbolic DAG onto vFPGA with pre-defined Python templates.}
    \vspace{-1em}
    \label{fig:dag}
\end{figure}

Figure~\ref{fig:dag} expands this view by showing the symbolic DAG that underlies compilation.
Here, stateless transforms are fused into a vectorized stage (Stage-A), while vocabulary operators introduce explicit broadcast and gather edges to model shared state access.
During the \emph{fit} phase, \textit{VocabGen} performs keyed reductions to construct the vocabulary table, which the compiler partitions across $P$ HBM banks for parallel access.
During the \emph{apply} phase, \textit{VocabMap} consumes this table with keyed lookups and produces transformed features that are gathered back into the output stream.
The mapping process instantiates user logic with fused pipelines, vocabulary operators, FIFOs, and broadcast/gather fabrics, while relying on the FPGA shell for DMA, RDMA offload, and memory/network arbitration.
In this way, the symbolic DAG is systematically lowered into a vFPGA implementation that couples operator fusion with partitioned state placement, enabling ETL at line rate.

\subsection{Operators in FPGA}

The process of feature engineering for recommender systems is an important aspect of ensuring optimal system performance. 
To accommodate both dense and sparse features with a unified interface, the adjustable data width is set to 64 bytes to match the data loading speed, enabling \textsc{PipeRec} to process multiple words in parallel.

\subsubsection{Dense Features}
Dense features contain important information across a continuous range. 
The industrial preprocessing pipeline includes operators to clip negative values to zero and apply a logarithmic transformation to reduce the skewness of data. 


\textit{Clamp}. 
This operation clips negative input dense features to zero while retaining positive values in their original form. 
\textsc{PipeRec} implements this operation using a ternary operator, achieving an Initiation Interval (II) of one cycle. 

\textit{Logarithm}. 
The logarithmic operation is effective in ML training, as it reduces skewness and manages extremely large values efficiently. 
\textsc{PipeRec} computes the logarithm using a hardware math library, achieving an II of one cycle. 

All operators achieve an initiation interval (II) of one cycle with minimal resource overhead.
By executing in a pipelined manner, the overall dataflow achieves an II of one cycle.

\subsubsection{Sparse Features}
The processing of sparse features involves key operations such as converting hexadecimal values to integers and applying a positive modulus to constrain the range of feature values. 
Besides these stateless operators, maintaining a vocabulary table with a stateful operator plays a crucial role by mapping columns of features to a range of continuous indices, which enables further operations like frequency-based filtering and initialization of lookups in a trainable embedding table.

\textit{Hex2Int}. 
Original categorical features are encoded as hexadecimal strings and it is necessary to convert them to integer values first. 
\textsc{PipeRec} performs this operation by translating each ASCII code into its binary representation and concatenating these to produce the result. 
This approach achieves an II of one cycle.

\textit{Modulus}. 
The modulus operation is designed to restrict the range of sparse feature values. 
\textsc{PipeRec} realizes this operation with the default math library, achieving an II of one cycle.

\textit{VocabGen}.
Categorical features represent attributes with specific, discrete values, and ML models perform optimally when trained on continuous values \cite{hancock2020survey}, making it crucial to create a tailored vocabulary table with unique indices for different columns \cite{meta_dlrm, tf_dlrm}. 
Given that dynamic vocabulary tables are frequently updated with new data, ensuring the efficiency of this step is vital for the end-to-end dataflow \cite{sima2022ekko}.  
\textsc{PipeRec} accelerates this process by constructing the vocabulary table in a pipelined manner. 
It efficiently processes streaming data from upstream modules, extracting unique values in a list, where the length of the list is determined by the range of \textit{Modulus}. 
The downstream module in \textsc{PipeRec} tracks the appearing sequence of occurrences for each unique value, assigns a corresponding index, and stores the value-index pair in memory.
The achieved II depends on the memory in use to store value-index pairs, with on-chip memory offering an II of two cycles due to Read-After-Write latency, and off-chip memory achieving an II of approximately six cycles.

\textit{VocabMap}.
The lookup process of mapping each value to its corresponding index makes \textsc{PipeRec} iterate over the entire dataset once more, and the final output comprises the corresponding indices in sequence.
\textsc{PipeRec} achieves this functionality using the pre-generated vocabulary table, reaching an II of approximately six cycles when the table is stored in off-chip memory or one cycle when stored in on-chip memory.  
This operator represents the final stage in handling sparse features, and II of the overall dataflow is influenced by the memory type used for vocabulary table.

\subsection{Data Loading and Memory Subsystem} 

\begin{figure}[t]
    \centering
    \includegraphics[width=0.6\linewidth]{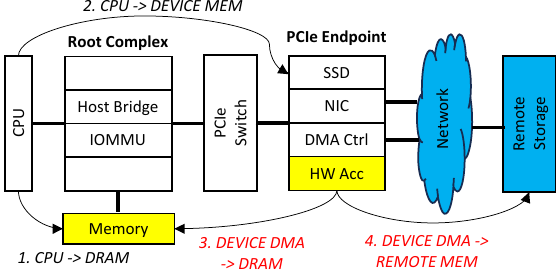}
    \caption{Data access pattern for PCIe-attached hardware accelerator in the host system.}
    \label{fig:data_access}
\end{figure}

Memory accesses play a critical role in the performance of ML systems, significantly affecting the speed, scalability, and efficiency of data processing. 
Efficient memory access contributes to minimize overhead, especially in scenarios involving large-scale data, real-time operations, and systems that demand high throughput and minimal latency.
However, understanding data access patterns is a non-trivial task. Figure \ref{fig:data_access} illustrates the data flow architecture utilized by \textsc{PipeRec}, showcasing a high-performance system where the CPU can offload compute-intensive tasks to specialized hardware while managing data transfers through PCIe and hierarchical memory structures. 
On the host side, the CPU can directly access DRAM or interact with PCIe-connected SSDs. 
PCIe-attached hardware accelerators, such as FPGAs or GPUs, can access local host memory through a DMA controller and extend their reach to remote memory via network interfaces. 
The integration of remote memory access enhances scalability, enabling utilization beyond local resources, which is particularly advantageous in cloud and data center environments. 
In the subsequent contents, we will demonstrate the distinct characteristics of accessing various types of memory.

\textbf{Device-attached Memory.}
Data-center FPGAs provide off-chip DDR/HBM in the tens-of-GB range \cite{alveo_u55c}. 
The many-channel organization of HBM delivers high aggregate bandwidth and is effective for DLRM-style workloads \cite{jiang2021microrec, zhu2021distributed}. 
Two limitations remain: capacity is orders of magnitude below TB–PB datasets, and non-pipelined host$\rightarrow$FPGA transfers are expensive—on our platform, transfer time is comparable to ETL execution.

\textbf{Host Memory.}
Over PCIe, the FPGA can stream directly from host DRAM via DMA engines (e.g., Xilinx DMA/Bridge for PCIe \cite{xilinx_dma} and Intel PCIe with DMA \cite{intel_dma}), exploiting servers’ hundreds of GB of memory with a simple streaming model. 
However, PCIe bandwidth (tens of GB/s) imposes a hard ceiling, and DRAM capacity is still finite; spilling to SSDs increases capacity at the cost of higher latency and lower bandwidth \cite{lee2024presto}.

\textbf{Remote Memory.}
Modern clusters increasingly disaggregate compute and storage. 
Accordingly, \textsc{PipeRec} accesses remote memory over RDMA using a custom hardware network stack \cite{sidler2020strom, he2024accl+}, bypassing TCP/IP overheads and sustaining high throughput. 
This approach introduces two primary constraints (see \S\ref{sec:resource_utilization}): (a) on-board resource contention under multi-tenant use, and (b) performance limited by available network bandwidth.

\begin{figure}[t]
    \centering
    \includegraphics[width=0.75\linewidth]{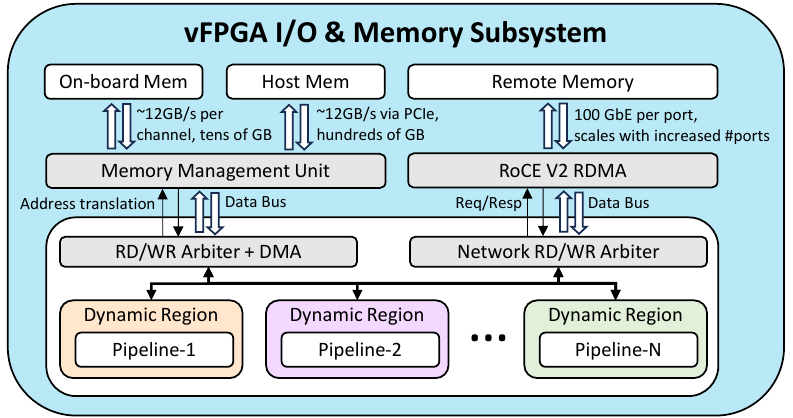}
    \caption{\textsc{PipeRec} vFPGA I/O \& memory subsystem with N reconfigurable pipelines. On-board memory, host memory (PCIe), and remote memory (RoCEv2) feed N pipelines via RD/WR crossbars. The MMU handles address translation; data moves over AXI buses.}
    \label{fig:memory_abstraction}
\end{figure}

Figure~\ref{fig:memory_abstraction} details the vFPGA I/O and memory subsystem that sustains the compiled dataflow.
Three memory classes feed the pipelines: on-board memory, host memory via PCIe, and remote memory via RoCEv2 RDMA.
A memory-management unit (MMU) performs address translation and exposes a unified virtual address space to the dataflow, decoupling operator logic from physical locations.
Read/write arbiters with DMA engines drive two crossbars: one for local/host memory traffic and one for network RDMA traffic, each exporting credit-based interfaces for backpressure.
Dynamic regions on the FPGA host $N$ reconfigurable pipelines; each region connects to the I/O subsystem through AXI streams and inherits the same virtual addressing model from the MMU.
Per-pipeline throughput is set by $(N \times W \times f_{\mathrm{clk}}) \times \mathrm{utilization}$ and is provisioned to match or exceed the downstream trainer’s consumption.
On-board and host memory deliver on the order of tens of GBps per channel, while network bandwidth scales with the number of 100 GbE ports; the arbiters and DMA engines multiplex these sources to keep the pipelines busy.
The resulting architecture allows state placement to be tuned (HBM for hot tables, BRAM for small metadata, host/remote memory for cold partitions) without changing operator code.
Together, the compiler and vFPGA subsystem map high-level ETL DAGs to deeply pipelined hardware that streams training-ready batches at line rate and exposes explicit backpressure to the GPU staging interface.

\subsection{Programmable Interface for Pipeline Composition}

Hardware accelerators are traditionally optimized for fixed designs, which limits their adaptability when serving diverse workloads. 
In practice, this raises two central challenges.
\textbf{Q1 (multi-tenancy)}: How can heterogeneous ETL pipelines coexist on the same FPGA without requiring costly bitstream recompilation? 
\textbf{Q2 (elasticity)}: How can pipeline performance be elastically scaled to match varying throughput and latency requirements?


To address these challenges, \textsc{PipeRec} exposes a \emph{Python-based template interface} for composing ETL pipelines. 
Instead of hand-crafting RTL or relying on black-box IPs, developers write pipelines in Python using pre-designed operators such as \texttt{Clamp}, \texttt{Modulus}, \texttt{VocabGen}, and \texttt{VocabMap}. 
A symbolic DAG is automatically constructed from the template, capturing both stateless operators (which can be fused into vectorized stages) and stateful operators (which require broadcast/gather access to shared tables in HBM). 
The DAG is then compiled into vFPGA dataflows with explicit operator fusion, table partitioning, and parallelism parameters ($N$, $W$). 
Before hardware mapping, \textsc{PipeRec} performs functional verification on the fused DAG to ensure operator semantics and schema constraints are preserved.  

Conventional methods such as packaging RTL IPs fall short for \textbf{Q1} and \textbf{Q2}, as they require full recompilation for each new workload. 
Instead, \textsc{PipeRec} employs partial reconfiguration: distinct ETL pipelines are loaded into dynamic regions at runtime, enabling concurrent execution of heterogeneous dataflows. 
As shown in Figure~\ref{fig:memory_abstraction}, each dynamic region hosts a pipeline instance with fine-grained columnar processing (e.g., 64-byte granularity). 
A unified I/O and memory subsystem provides scalable access to on-board HBM/DDR, PCIe-attached host memory, and RDMA-enabled remote memory, coordinated via translation lookaside buffers (TLBs), arbiters, and AXI interconnects.  
This design ensures that multiple pipelines can safely share resources while dynamically adapting performance to workload demands.  

Together, the Python template interface for DAG compilation (Figure~\ref{fig:dag}) and the reconfigurable I/O and memory subsystem (Figure~\ref{fig:memory_abstraction}) enable \textsc{PipeRec} to meet both multi-tenancy and elasticity requirements, delivering a practical programmable interface for FPGA-based ETL in continuous training pipelines.

\subsection{FPGA--GPU Streaming Dataflow}\label{sec:cpu-vs-fpga}

\begin{figure}[t]
    \centering
    \begin{subfigure}[b]{0.7\textwidth}
        \includegraphics[width=1\linewidth]{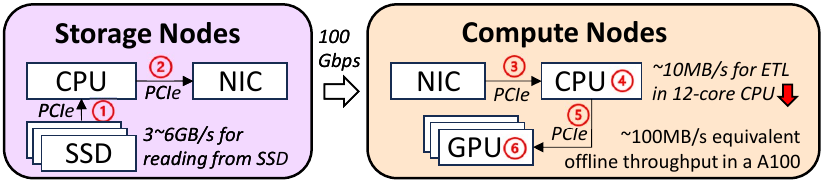}
        \caption{Traditional CPU-based ETL for  online recommender model training, where the bottleneck is the preprocessing capability in the CPU.}
        \label{fig:cloud_cpu}
    \end{subfigure}
    \hfill 
    \begin{subfigure}[b]{0.7\textwidth}
        \includegraphics[width=1\linewidth]{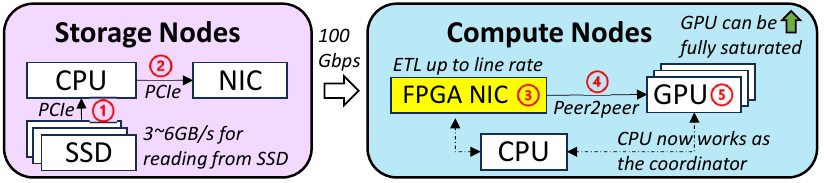}
        \caption{FPGA-accelerated ETL in \textsc{PipeRec}, where FPGA-GPU co-design makes GPU fully utilized.}
        \label{fig:cloud_piperec}
    \end{subfigure}

    \caption{
        CPU-based vs. FPGA-accelerated ETL pipelines for online recommender training.
        (a) In CPU-based ETL, preprocessing ($\sim$10 MB/s in a 12-core CPU) lags behind GPU training ($\sim$100 MB/s), leaving GPUs underutilized.
        (b) In \textsc{PipeRec}, an FPGA-NIC performs ETL at line rate and streams data directly to GPUs via peer-to-peer transfer, while the CPU only coordinates. This removes the CPU bottleneck and enables full GPU utilization.
        }
    \label{fig:cpu-vs-fpga-etl}
\end{figure}

Figure~\ref{fig:cpu-vs-fpga-etl} contrasts a traditional CPU-based pipeline with the proposed FPGA-accelerated approach.
In the CPU-based design (Figure~\ref{fig:cloud_cpu}), data flow from storage to compute nodes over a 100~Gbps fabric, but preprocessing on a 12-core CPU sustains only on the order of $\sim$10~MB/s (values extracted from Figure~\ref{fig:preprocess_train}).
GPU training can consume data at roughly an order of magnitude higher rate ($\sim$100~MB/s), so the ETL stage dominates wall-clock time and leaves accelerators underutilized.
Additional copies through host memory and PCIe further widen the gap between ingestion and training.

In the FPGA-accelerated design (Figure~\ref{fig:cloud_piperec}), ETL operators run on a NIC-attached or PCIe-attached FPGA and stream batches directly into GPU memory via peer-to-peer transfer.
The CPU acts as a coordinator for control metadata, while the FPGA and GPU form a streaming data path with explicit credits for rate matching.
By removing contention for GPU cores and HBM, the design sustains line-rate ingestion on the ETL side and keeps the trainer continuously fed.
Because ETL demand scales with data volume rather than model size, ETL can be sharded across FPGAs independently of the number of trainers, while GPUs are reserved for forward/backward passes and embedding lookups.
The result is a balanced pipeline that avoids GPU idling, reduces time-to-freshness, and preserves programmability through a high-level ETL abstraction compiled to hardware dataflows.

\section{Evaluation}

We evaluate \textsc{PipeRec} through the following questions:
\begin{itemize}
    \item [\URoman{1}.] Which operators benefit most from hardware acceleration? How does performance differ between stateless and stateful operators? $\S$\ref{sec:micro-benchmark}
    \item [\URoman{2}.] What performance gains does \textsc{PipeRec} achieve over server-grade CPUs and modern GPUs? $\S$\ref{sec:stateless}, $\S$\ref{sec:stateful}
    \item [\URoman{3}.] How does the power efficiency of \textsc{PipeRec} compare to other baselines? $\S$\ref{sec:power_efficiency}
    \item [\URoman{4}.] How can \textsc{PipeRec} scale throughput by supporting multiple concurrent dataflows? $\S$\ref{sec:scalability}
\end{itemize}

\subsection{Experimental Setup}

\subsubsection{Dataset}
We evaluate the effectiveness of \textsc{PipeRec} on both real and synthetic datasets.  
For the real case, we leverage the Criteo Kaggle dataset (\textbf{Dataset-I})~\cite{criteo}, a well-known public dataset for recommender systems, containing multi-day online advertising data.  
The dataset is originally stored in UTF-8 format, which introduces additional overhead due to decoding and is targeted for row-based processing.  
To enhance performance, we extract binary data for memory alignment and verify the data format of each feature.  
For efficient columnar processing and to focus on the capabilities of preprocessing pipelines, we store the binary data as a Parquet file without compression for all solutions.  
We compare the speed of data loading and preprocessing to ensure that the data loading process does not become a bottleneck.  
In this transformation, dense features are represented as floating-point values, and sparse features are represented as fixed-length hexadecimal strings.  
The size of the transformed dataset is 17GB and it includes 45 million entries, consisting of 13 dense features and 26 sparse features.  
Furthermore, we create a synthetic dataset (\textbf{Dataset-II}) with 4 million entries to facilitate additional analysis.  
This synthetic dataset is expanded to include 504 dense features and 42 sparse features, with a size of 11GB~\cite{zhao2022understanding, lee2024presto}.  
In addition, we include the Criteo 1TB click logs dataset (\textbf{Dataset-III})~\cite{criteo_1T} to emulate industrial-scale workloads and evaluate ETL performance under large-scale data ingestion scenarios.
This dataset is sharded into 1024 Parquet files to increase the I/O performance and the total size is about 1.5TB.

\subsubsection{Hardware Platform}
We evaluate three classes of platforms: CPUs, GPUs, and FPGAs.  
(1) For CPU baselines, we use Apache Beam on Google Cloud N2 instances (n2-standard-16/32/64/96/128), as well as a server-grade two-socket machine equipped with AMD EPYC 7V13 (128 cores, 512 GB DRAM).  
(2) For GPU experiments, we use both cloud and local setups: a Google Cloud instance with an Nvidia A100 (40 GB HBM, 12-core CPU, 85 GB RAM, 2 TB balanced persistent disk, and 4$\times$375 GB local NVMe SSDs), and a local workstation with an RTX 3090 (24 GB GDDR6X, 64-core CPU, 252 GB RAM, 2 TB NVMe SSD).  
(3) For \textsc{PipeRec}, we use a Xilinx Alveo U55c with 16 GB HBM (32 channels, peak throughput 460 GB/s) and 43 MB SRAM. The host system has an AMD EPYC 7302P 16-core CPU, and an AMD MI210 GPU is attached for end-to-end benchmark. Unless otherwise specified, the achieved kernel frequency of \textsc{PipeRec} is 200 MHz.

\subsubsection{Experiment Configuration}\label{sec:exp_config}
\begin{figure}[t]
    \centering
    \includegraphics[width=0.65\linewidth]{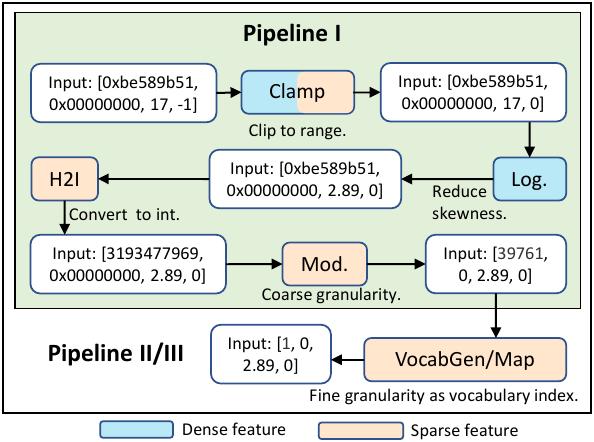}
    \caption{Representation of pipelines in use for evaluation (Log = Logarithm, H2I = Hex2Int, Mod = Modulus). The input values include two sparse features and two dense features.}
    \label{fig:pipelines}
\end{figure}

We evaluate three representative preprocessing pipelines for recommender systems across three datasets. 
These pipelines cover different configurations: \textit{a stateless pipeline}, \textit{a stateful pipeline with low memory intensity}, and \textit{a stateful pipeline with high memory intensity}. Figure \ref{fig:pipelines} shows a concrete example of data transformation.

\begin{itemize}  
    \item \textbf{Pipeline I}: A stateless pipeline that processes both dense and sparse features using operators such as \textit{Clamp, Logarithm, Hex2Int}, and \textit{Modulus} (Table~\ref{tab:operator_list}).  
    \item \textbf{Pipeline II}: A stateful pipeline that extends Pipeline~I with additional small vocabulary tables for sparse features.  
    \item \textbf{Pipeline III}: A stateful pipeline that extends Pipeline~I with large vocabulary tables for sparse features.  
\end{itemize}

\subsubsection{Implementation Details}
We use the Google Cloud Dataflow and Apache Beam with varying numbers of machines as the default baseline, which is used in Google's solution \cite{tf_dlrm}, focusing on processing large-scale data in distributed environments.
To better understand the processing capability of CPU, we rely on \textit{pandas, numpy, joblib} to run columnar preprocessing tasks for Parquet in parallel in server-grade CPU, and use \textit{s-tui} \cite{s-tui} to monitor its power consumption.
For the execution in GPU, we run with Nvidia Merlin NVTabular and RAPIDS \textit{dask-cudf}, and use \textit{nvidia-smi} to record the power and resource utilization accordingly.
 
We use Vitis HLS \cite{vitis_hls}, a high-level synthesis tool for Verilog, to implement the code for the memory interface and the preprocessing operator in \textsc{PipeRec}. 
We use Xilinx XRT \cite{xilinx_xrt} to verify the functionality of hardware codes and convert to IP blocks for the next usage.
We leverage an open-source FPGA shell, Coyote \cite{korolija2020abstractions}, to facilitate dynamic regions and partial reconfiguration with manual floorplanning. 
These dynamic regions empower clients to execute diverse preprocessing pipelines concurrently, and \textsc{PipeRec} can complete the transition of pipelines within milliseconds. 
For RDMA communication, we rely on open-source hardware modules \cite{sidler2020strom}, which is written in Verilog and integrates seamlessly with Coyote.

To illustrate the dynamic scalability, we use Pipeline I (shown in $\S$ \ref{sec:exp_config}) to show how to operate several pipelines concurrently. For this purpose, we integrate the generated IP blocks into Coyote, configure the memory interface and dynamic regions, and then complete the compilation flow. 
Based on an open source duplex RoCE v2 RDMA network protocol\cite{sidler2020strom}, \textsc{PipeRec} directly accesses data from the remote node and processes it over RDMA data streams. 
The configured dynamic region efficiently routes incoming data from the network stack, executes the preprocessing pipeline, and writes the processed data to the target position.
Table \ref{tab:resource_fpga} displays the resource utilization for three preprocessing pipelines (local and remote) and the pure RDMA hardware stack.
Here we focus on the utilization of CLB, BRAM and DSP while neglecting unused URAM and other resources.

\subsection{Evaluation Baselines}

\subsubsection{Disadvantage of Von-Neumann Architecture}
Unlike CPUs and GPUs, which follow the von Neumann model with fixed instruction pipelines, FPGAs are reconfigurable fabrics that directly realize dataflow-style execution. 
On CPUs and GPUs, ETL pipelines are executed as a sequence of kernels, with each operator materializing its output in global memory before the next can proceed—a pattern that incurs high latency and energy due to excessive memory traffic. 
By contrast, FPGA pipelines connect operators through on-chip FIFOs and registers, eliminating intermediate materialization and enabling streaming execution with fine-grained parallelism.

\subsubsection{CPU Baseline}\label{sec:cpu_baseline}
We select two widely used CPU-based software frameworks as baselines: Apache Beam and Pandas.  
Apache Beam, executed remotely on Google Cloud Dataflow, represents a distributed ETL framework designed for scalability. Following the setup in the tutorial \cite{tf_dlrm}, we first convert and shard the raw dataset into multiple Parquet files to increase data loading throughput from Google Cloud Bucket (approximately 700~MB/s within the same region), and finish experiments in sequence.  

\begin{figure}[t]
    \centering 
    \includegraphics[width=0.8\linewidth]{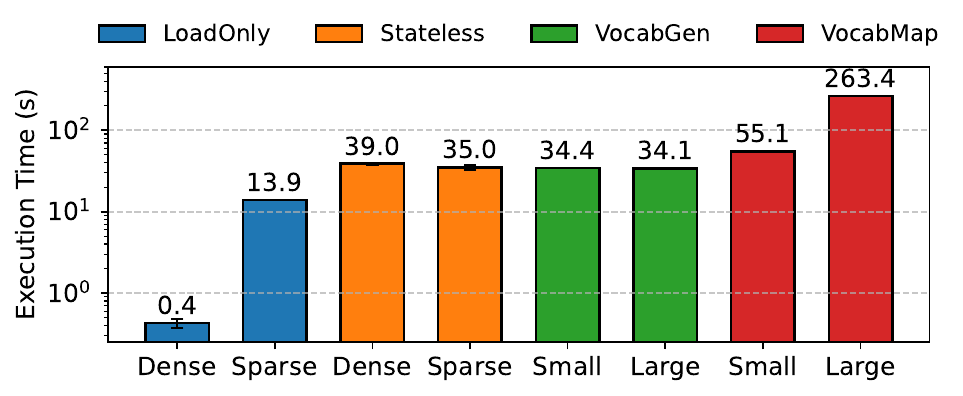}
    \caption{Micro-benchmark II: execution time of preprocessing pipelines for a single feature using one thread. X-axis labels denote feature types (Dense, Sparse, Small, Large), while colors group operations into four pipelines: (i) LoadOnly — loading from memory; (ii) Stateless — Clamp and Logarithm (dense) or Hex2Int and Modulus (sparse); (iii) VocabGen — generating vocabulary tables (small/large) for sparse features; (iv) VocabMap — mapping sparse features using vocabulary tables (small/large).
    }
    \label{fig:pipeline_independent}
\end{figure}

In contrast, Pandas serves as a local, single-node baseline with optimized vectorized operations.  
To ensure fairness, we adopt several best practices: warming up to preload datasets into CPU memory (for Dataset~I and II) or NVMe SSD (for Dataset~III), repeating experiments to report average performance with variance, and decomposing pipeline stages (Figure~\ref{fig:pipeline_independent}) to identify bottlenecks.  
Our analysis shows that dense feature loading is efficient, while sparse feature preprocessing---particularly vocabulary table construction and lookups---dominates runtime.  
This confirms that preprocessing computations, rather than data loading, are the main performance bottleneck in CPU-based pipelines.  

\begin{figure}[t]
    \centering
    \begin{subfigure}[b]{1\textwidth} 
        \centering 
        \includegraphics[width=0.6\linewidth]{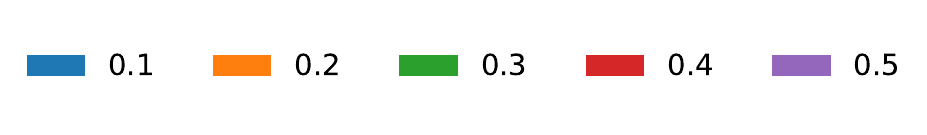}
        \vspace{-1em}
    \end{subfigure}
    \begin{subfigure}[b]{0.45 \textwidth} 
        \centering 
        \includegraphics[width=0.8\linewidth]{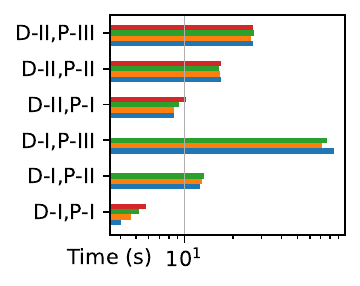}
        \caption{RTX 3090}
        \label{fig:gpu_3090}
    \end{subfigure}
    \begin{subfigure}[b]{0.45\textwidth} 
        \centering 
        \includegraphics[width=0.8\linewidth]{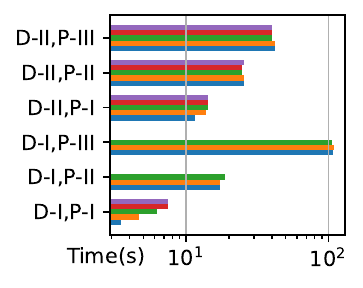}
        \caption{A100}
        \label{fig:gpu_a100}
    \end{subfigure}
    \caption{Impact of GPU memory fractions for various preprocessing configurations. Dataset I+Pipeline-I is abbreviated as D-I,P-I, and the same for others.}
    \vspace{-1em}
    \label{fig:gpu_performance}
\end{figure}

\begin{figure}[t]
    \centering
    \begin{subfigure}[b]{1\textwidth} 
        \centering 
        \includegraphics[width=0.6\linewidth]{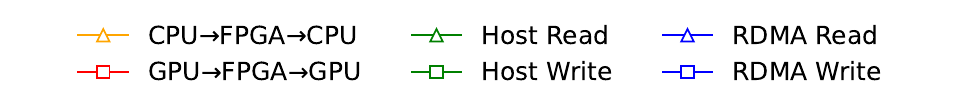}
    \end{subfigure}
    \vspace{-1em}
    \begin{subfigure}[b]{0.45\textwidth} 
        \centering 
        \includegraphics[width=0.95\linewidth]{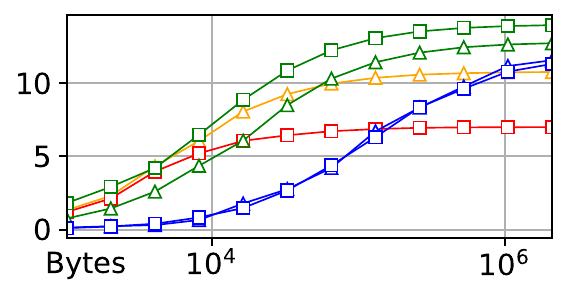}
        \caption{Throughput (GBps)}
        \label{fig:host_rdma_throughput}
    \end{subfigure}
    \hfill 
    \begin{subfigure}[b]{0.45\textwidth} 
        \centering 
        \includegraphics[width=1\linewidth]{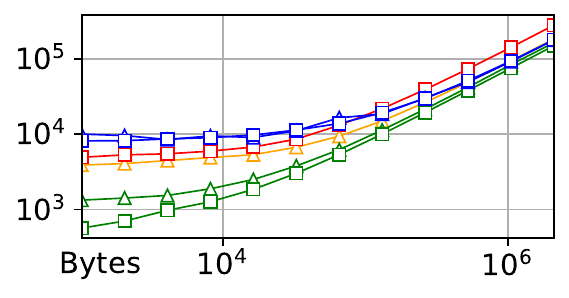}
        \caption{Latency (ns)}
        \label{fig:host_rdma_latency}
    \end{subfigure}

    \caption{Micro-benchmark I: measurement of throughput and latency for data movement in \textsc{PipeRec}. The command for host/RDMA read/write is initiated from vFPGA.}
    \label{fig:host_rdma}
\end{figure}

\subsubsection{GPU Baseline}
We also benchmark NVIDIA NVTabular as a representative GPU baseline \cite{nvtabular_blog, nvidia_nvtabular_dlrm}.  
NVTabular is designed for preprocessing large-scale tabular datasets, leveraging GPU parallelism, memory bandwidth, and optimized kernels for both dense and sparse transformations (e.g., clipping, modulus, and categorization).  
Its columnar processing model minimizes data movement and enables concurrent execution across multiple features.  

To handle datasets larger than GPU memory, we configure a Dask-based streaming pipeline, where data is partitioned into manageable chunks (e.g., 1~GB) and processed with NVTabular using the RAPIDS Memory Manager (RMM) pool.  
As shown in Figure~\ref{fig:gpu_performance}, increasing the GPU RMM pool fraction from 0.1 to 0.5 affects NVTabular runtime across all pipelines, with most of the gain realized by $\sim$0.3 and only modest improvements thereafter on both RTX~3090 and A100.
This out-of-core configuration enables NVTabular to efficiently scale to multi-terabyte datasets.  
Prior reports \cite{nvtabular_blog} demonstrate significant acceleration for Criteo ETL workloads, making it a strong GPU baseline for our evaluation.  

\subsection{Micro-benchmarks}\label{sec:micro-benchmark}
We conduct three micro-benchmarks to understand the performance of ETL pipelines.  

Figure~\ref{fig:host_rdma} reports throughput and latency versus transfer size for host--FPGA DMA (read/write), the end-to-end CPU$\rightarrow$FPGA$\rightarrow$CPU and GPU$\rightarrow$FPGA$\rightarrow$GPU paths, and RoCEv2 RDMA. 
Throughput increases with message size and plateaus beyond $\sim$1\,MiB: host DMA peaks at approximately 12--14~GB/s, the CPU$\rightarrow$FPGA$\rightarrow$CPU path reaches approximately 12--13~GB/s, the GPU$\rightarrow$FPGA$\rightarrow$GPU path saturates near 7~GB/s, and RDMA sustains approximately 11--12~GB/s (close to 100\,GbE line rate). 
Latency exhibits the complementary trend: small transfers are dominated by setup costs (host: $\sim$0.6--1.5\,$\mu$s; RDMA: $\sim$8--10\,$\mu$s) and then grow roughly linearly with payload size. 
These observations motivate batching into MiB-scale chunks and overlapping transfer with compute (e.g., double-buffered DMA/RDMA) to keep pipelines saturated.

Figure~\ref{fig:pipeline_independent} reports the execution time of preprocessing pipelines for a single feature using one CPU thread.  
We compare four categories of pipelines: (i) \textbf{LoadOnly}, which measures the baseline cost of loading from memory; (ii) \textbf{Stateless}, which applies simple transformations such as \textit{Clamp} and \textit{Logarithm} for dense features or \textit{Hex2Int} and \textit{Modulus} for sparse features; (iii) \textbf{VocabGen}, which generates vocabulary tables of varying sizes; and (iv) \textbf{VocabMap}, which maps sparse features using the generated vocabulary tables.  
We observe that \textbf{LoadOnly} incurs negligible cost, while stateless operators add moderate overhead.  
Vocabulary-related pipelines dominate execution time, especially \textbf{VocabMap} with large tables, which becomes the primary bottleneck.  

Table~\ref{tab:operators} further compares operator execution times on different platforms, including CPU, RTX~3090, A100, and \textsc{PipeRec}.  
On CPUs, lightweight operations such as Logarithm, Hex2Int, and Modulus require hundreds of seconds, and large vocabulary operations are prohibitively slow (e.g., 2390 seconds for VocabMap-512K).  
In contrast, GPUs achieve several orders of magnitude lower latency for stateless and mapping operators, although vocabulary generation remains costly (e.g., $\sim$64--69 seconds for 512K).  
\textsc{PipeRec} delivers competitive performance across all operators, particularly reducing the cost of large vocabulary operations by more than two orders of magnitude compared to CPUs.  
These results highlight that while GPUs excel at lightweight transformations, \textsc{PipeRec} provides balanced and efficient acceleration across both stateless and vocabulary-heavy pipelines.

\begin{table}[t]

    \caption{Micro-benchmark III: per-operator runtime on Dataset I across platforms (seconds).}
    \label{tab:operators}
    \centering
    \scalebox{0.85}{
    \begin{tblr}{
      colspec={ccccc},
    }
    \hline
    \textbf{Operators} & \textbf{CPU} & \textbf{RTX 3090} & \textbf{A100} & \textbf{\textsc{PipeRec}}\\
    \hline
    Clamp & 4.20$\pm$0.01 & 0.029$\pm$3E-4 & 0.043$\pm$5E-4 & 0.23 \\
    Logarithm & 475.28$\pm$2.31 & 0.01$\pm$4E-5 & 0.015$\pm$1E-4 & 0.23 \\
    Hex2Int & 410.59$\pm$10.45& 0.051$\pm$2E-3 & 0.059$\pm$1E-4 & 0.92 \\
    Modulus & 354.25$\pm$2.42 & 0.017$\pm$2E-4 & 0.026$\pm$1E-4 & 0.46 \\
    VocabGen-8K & 4.97$\pm$0.17 & 7.57$\pm$0.17 & 8.76$\pm$0.037 & 0.92 \\
    VocabMap-8K & 21.94$\pm$0.13 & 0.02$\pm$1E-3 & 0.11$\pm$1E-3 & 0.46 \\
    VocabGen-512K & 549.79$\pm$10.20 & 64.10$\pm$2.55 & 69.03$\pm$0.23 & 2.15 \\
    VocabMap-512K & 2390.26$\pm$10.26 & 0.015$\pm$1E-3 & 0.11$\pm$1E-3 & 2.96 \\

    \hline
    \end{tblr}
    }
    \begin{minipage}{0.95\linewidth}
        \footnotesize \textsuperscript{\dag}\,In NVTabular, vocabulary construction and lookup run as a combined \emph{fit–apply} workflow rather than independent operators. We report them separately here for consistency with CPU and \textsc{PipeRec}.
    \end{minipage}
\end{table}

\subsection{Stateless ETL Tasks}\label{sec:stateless}

\begin{figure}[t]
    \centering
    \begin{subfigure}[b]{0.75\textwidth}
        \centering 
        \hspace*{2em}
        \includegraphics[width=0.85\linewidth]{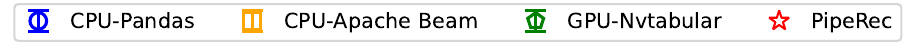}
    \end{subfigure}
    \begin{subfigure}[b]{0.75\textwidth}
        \centering 
        \includegraphics[width=0.95\linewidth]{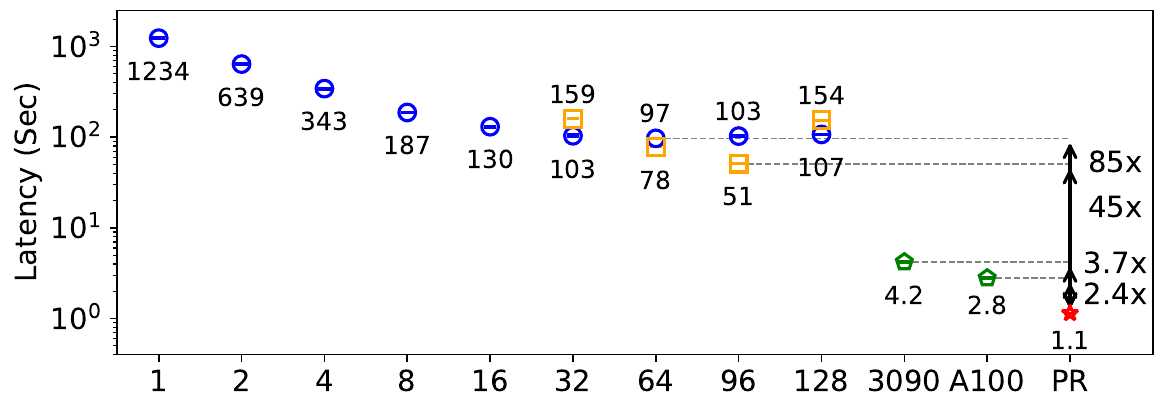}
        \caption{Dataset I + Pipeline I. PR represents \textsc{PipeRec}}
        \label{fig:p1_d1}
    \end{subfigure}
    \hfill 
    \begin{subfigure}[b]{0.75\textwidth}
        \centering 
        \includegraphics[width=0.95\linewidth]{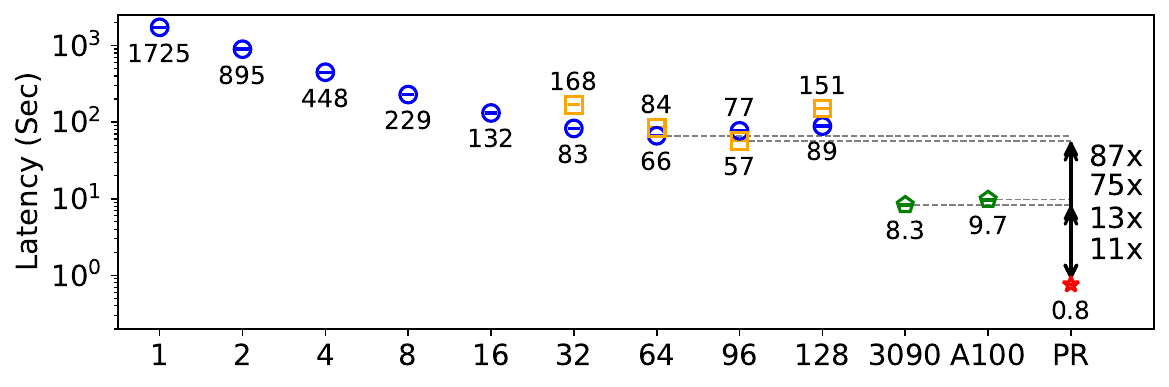}
        \caption{Dataset II + Pipeline I.}
        \label{fig:p1_d2}
    \end{subfigure}
    \hfill 
    \begin{subfigure}[b]{0.75\textwidth}
        \centering 
        \includegraphics[width=1\linewidth]{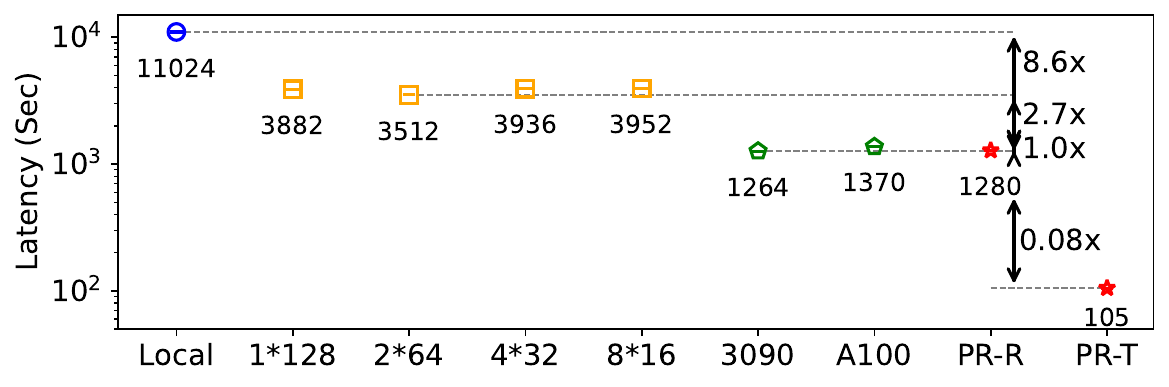}
        \caption{Dataset III + Pipeline I. 
            Local (Pandas): best run on a 128-core CPU, achieved with 64 threads.
            Apache Beam: cluster sized to 128 vCPUs total.
            Both GPU and \textsc{PipeRec} (PR-R) are bound by SSD reads ($\sim$1.2\,GB/s). 
            \textbf{PR-T} shows the theoretical lower bound of \textsc{PipeRec} without I/O limit.
            }
        \label{fig:p1_d3}
    \end{subfigure}
    \caption{
        Latency comparison across platforms for Pipeline I.
        Lower is better.
        }
    \label{fig:p1}
\end{figure}

We begin by analyzing the performance of stateless ETL pipelines.  
Figure~\ref{fig:p1} compares the latency of Pipeline~I across three datasets on different platforms.  
Pipeline~I consists solely of stateless operators, including \textit{Clamp, Logarithm, Hex2Int}, and \textit{Modulus}.  
CPU-based Pandas consistently exhibits the longest latency, even when parallelized on multi-core processors.  
Apache Beam provides scalability through distributed execution, but its benefit diminishes with larger cluster sizes due to coordination overhead.  
GPU acceleration with NVTabular substantially reduces runtime, achieving up to $3.7\times$ speedup over optimized CPU baselines.  
In contrast, \textsc{PipeRec} achieves the lowest latency across all datasets.  
On Dataset~I and II, \textsc{PipeRec} outperforms Pandas by $85\times$ and $87\times$, respectively.  
For Dataset~III (same column structure as Dataset~I), both the GPU baseline and \textsc{PipeRec} are \emph{SSD-bound}: throughput is limited by the $\sim$1.2\,GB/s read bandwidth (PR-R, read-bound). 
Considering that \textsc{PipeRec} provides a deterministic ETL process, the point at 105\,s (PR-T, theoretical) is the theoretical lower bound of \textsc{PipeRec} without I/O limit. 
In contrast, Dataset~I removes the I/O bottleneck and compares compute-bound performance directly between the GPU framework and \textsc{PipeRec}. 

\begin{figure}[b]
    \centering 
    \includegraphics[width=0.75\linewidth]{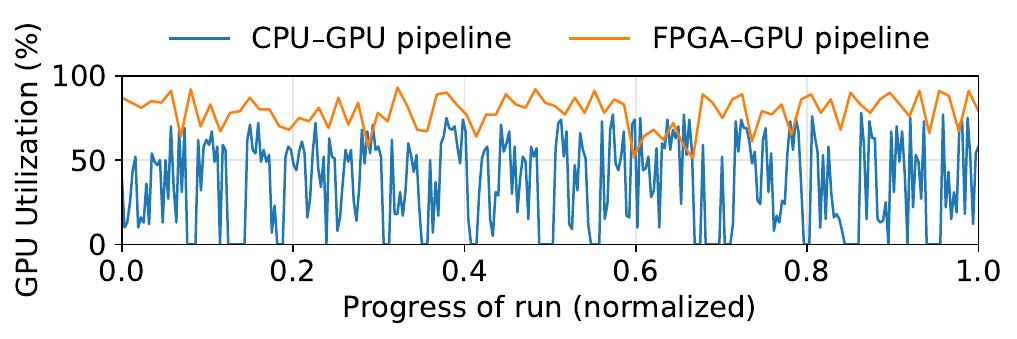}
    \caption{Normalized GPU utilization during training.
    }
    \label{fig:gpu_utilization}
\end{figure} 

\begin{figure*}[t]
    \centering
    \begin{subfigure}[b]{0.6\textwidth}
        \centering 
        \includegraphics[width=1\linewidth]{figures/pipeline_no_vocab_legend.pdf}
        \vspace{-1em} 
        \label{fig:pipeline_no_vocab_legend}
    \end{subfigure}
    
    \begin{subfigure}[b]{0.75\textwidth}
        \centering 
        \includegraphics[width=1\linewidth]{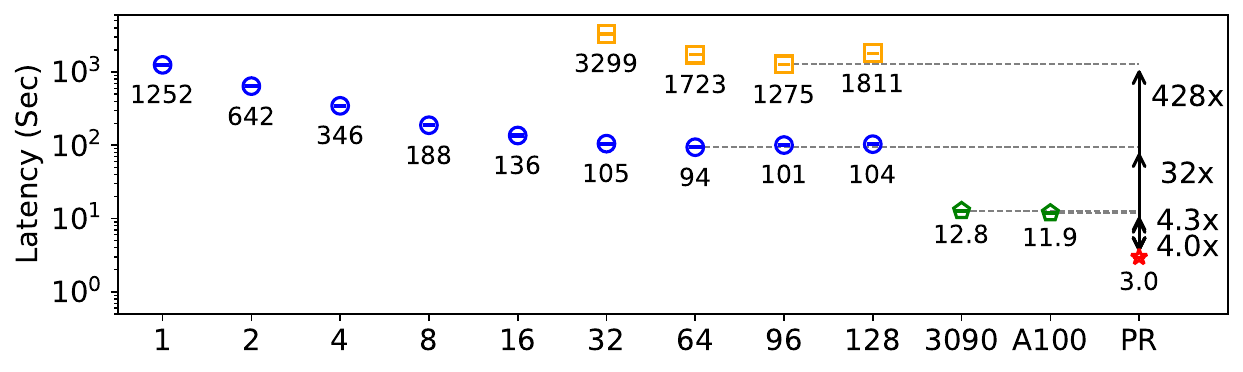}
        \caption{Dataset \URoman{1} + Pipeline \URoman{2}.}
        \label{fig:p2_d1}
    \end{subfigure}

    \begin{subfigure}[b]{0.75\textwidth}
        \centering 
        \includegraphics[width=1\linewidth]{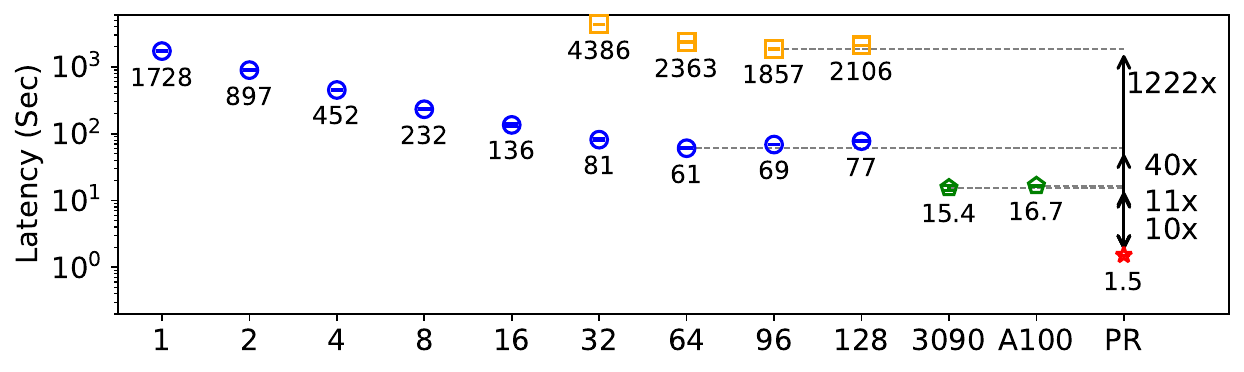}
        \caption{Dataset \URoman{2} + Pipeline \URoman{2}.}
        \label{fig:p2_d2}
    \end{subfigure}
    \begin{subfigure}[b]{0.75\textwidth}
        \centering 
        \includegraphics[width=1\linewidth]{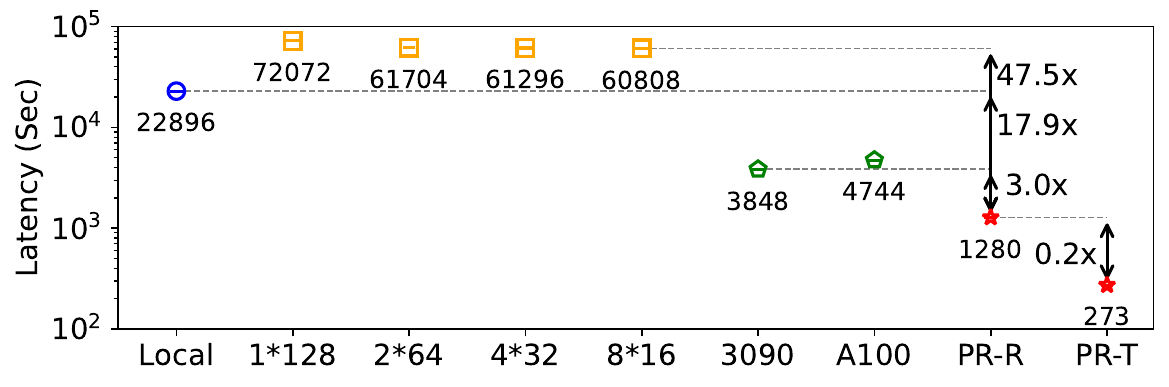}
        \caption{Dataset \URoman{3} + Pipeline \URoman{2}.}
        \label{fig:p2_d3}
    \end{subfigure}
    \caption{Latency comparison of ETL performance across platforms for stateful \textbf{Pipeline II}. For Dataset~III, \textsc{PipeRec} is limited by SSD read bandwidth, whereas the GPU baseline is not (compute-bound).
Lower is better.}
    \vspace{-1em}
    \label{fig:pipeline_all_p2}
\end{figure*}

\begin{figure*}[t]
    \centering
    \begin{subfigure}[b]{0.6\textwidth}
        \centering 
        \includegraphics[width=1\linewidth]{figures/pipeline_no_vocab_legend.pdf}
        \vspace{-1em} 
        \label{fig:pipeline_no_vocab_legend}
    \end{subfigure}
    \begin{subfigure}[b]{0.75\textwidth}
        \centering 
        \includegraphics[width=1\linewidth]{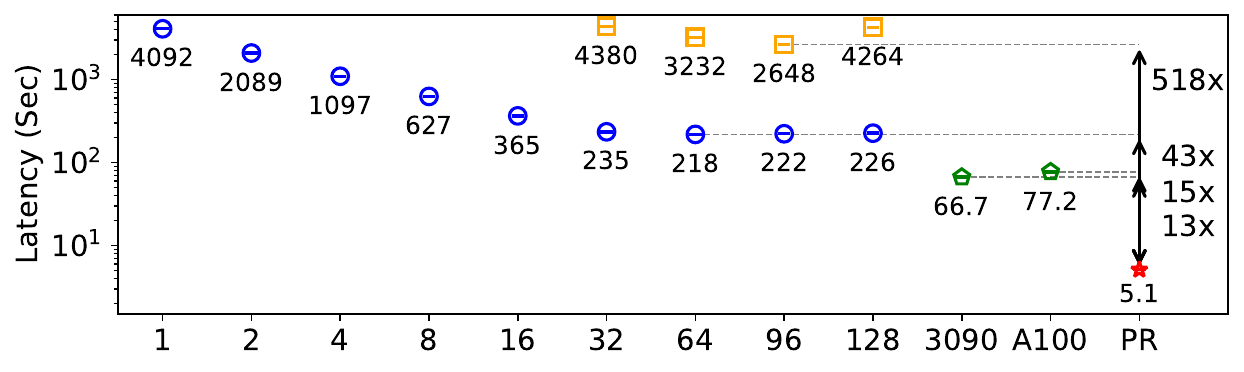}
        \caption{Dataset \URoman{1} + Pipeline \URoman{3}.}
        \label{fig:p3_d1}
    \end{subfigure}

    \begin{subfigure}[b]{0.75\textwidth}
        \centering 
        \includegraphics[width=1\linewidth]{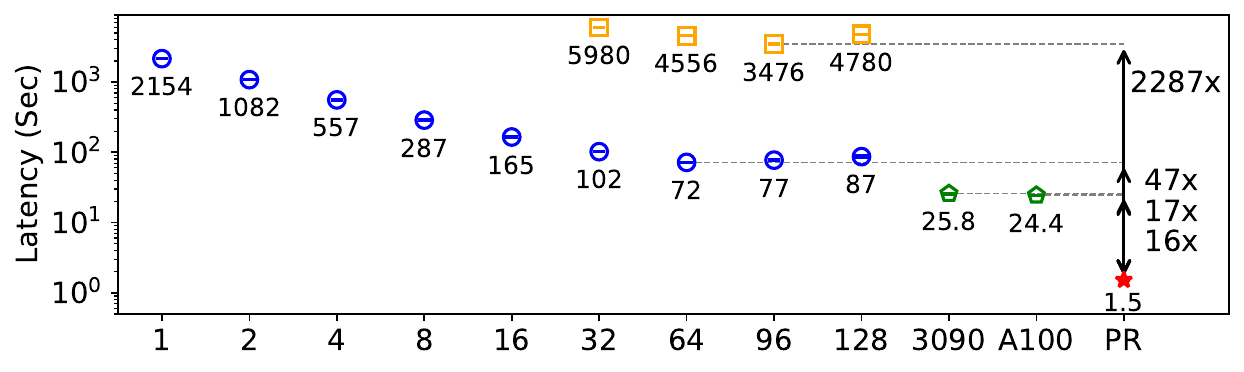}
        \caption{Dataset \URoman{2} + Pipeline \URoman{3}.}
        \label{fig:p3_d2}
    \end{subfigure}
    \begin{subfigure}[b]{0.75\textwidth}
        \centering 
        \includegraphics[width=1\linewidth]{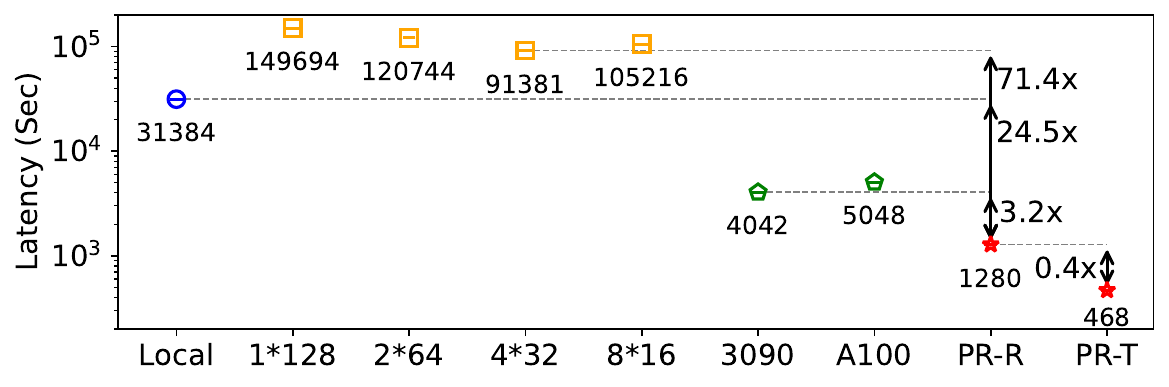}
        \caption{Dataset \URoman{3} + Pipeline \URoman{3}.}
        \label{fig:p3_d3}
    \end{subfigure}
    \caption{Latency comparison of ETL performance across platforms for stateful \textbf{Pipeline III}, which involves a higher number of random memory access. For Dataset~III, \textsc{PipeRec} is limited by SSD read bandwidth, whereas the GPU baseline is not (compute-bound).
Lower is better.}
    \vspace{-1em}
    \label{fig:pipeline_all_p3}
\end{figure*}

Figure~\ref{fig:gpu_utilization} shows the normalized GPU utilization during end-to-end training.  
In the CPU–GPU pipeline, irregular data delivery from CPU-based preprocessing leads to frequent stalls, causing highly unstable utilization that fluctuates between 0\% and 80\%.  
By contrast, the FPGA–GPU pipeline in \textsc{PipeRec} ensures stable and near-saturated GPU utilization throughout the run.  
This demonstrates that accelerating stateless ETL not only improves preprocessing latency but also eliminates bottlenecks in feeding data to GPUs, thereby maximizing overall training efficiency.  

\subsection{Stateful ETL Tasks}\label{sec:stateful}
We next evaluate stateful ETL tasks, which involve vocabulary operations that require maintaining state across samples.  
Pipeline~II corresponds to preprocessing with a small vocabulary table, while Pipeline~III uses a large vocabulary table.  
Figure~\ref{fig:pipeline_all_p2} and \ref{fig:pipeline_all_p3} presents the latency results across three datasets and four platforms: CPU-Pandas, CPU-Apache Beam, GPU-NVTabular, and \textsc{PipeRec}.  

For Dataset~I and II, CPU-based solutions again show the highest latency.  
While Apache Beam provides some convenience via distributed execution, the benefit remains limited due to synchronization and communication costs.  
GPU acceleration reduces latency by one order of magnitude compared to CPUs, but the performance gap widens as vocabulary size grows.  
Specifically, the latency of GPU-based NVTabular increases substantially when scaling from Pipeline~II to Pipeline~III, highlighting the high cost of large vocabulary operations.  
In contrast, \textsc{PipeRec} consistently achieves the lowest latency across all cases.  
On Dataset~I, \textsc{PipeRec} improves over Pandas by up to $32\times$ for Pipeline~II and $43\times$ for Pipeline~III.  
For Dataset~II, the corresponding improvements are $40\times$ and $47\times$.  
On Dataset~III, where the vocabulary table size dominates, \textsc{PipeRec} completes in $1280$ seconds for Pipeline~II and Pipeline~III, approaching the lower bound imposed by data loading throughput and indicating a significantly higher theoretical processing capability.  
Compared to GPU-NVTabular, these represent $3\times$–$17\times$ speedups depending on the dataset and vocabulary size.  
These results highlight that stateful vocabulary operations are the primary bottleneck in large-scale ETL pipelines.   \textsc{PipeRec} effectively mitigates this bottleneck and sustains performance even with high-cardinality vocabulary tables.

\subsection{Power Efficiency across Platforms}\label{sec:power_efficiency}

We now analyze the power efficiency of different platforms.  
Table~\ref{tab:power} reports both the average dynamic power consumption and the corresponding latency across datasets and pipelines.  
The static powers of each platform are 150W for CPU, 33W for RTX~3090, 43W for A100, and 17W for \textsc{PipeRec}.  
To quantify energy efficiency, we compute performance-per-watt (Perf/W), defined as the reciprocal of the product of latency and power.  
Since all platforms process the same dataset, we normalize Perf/W relative to the CPU baseline.  
The results show that CPUs consistently consume the most power (294--379W) yet deliver the lowest efficiency, which we set as $1.0\times$.  
GPUs improve efficiency by up to two orders of magnitude in lightweight pipelines, but their advantage diminishes with larger vocabulary sizes due to increased latency and power draw.  
For example, in D-I + P-I, the RTX~3090 and A100 achieve $59.4\times$ and $107.8\times$ higher Perf/W than the CPU, respectively, but in D-I + P-III the efficiencies drop to only $7.15\times$ and $11.3\times$.  
In contrast, \textsc{PipeRec} sustains consistently low power consumption (24--26W) across all configurations and achieves up to $1101.4\times$ higher Perf/W in D-II + P-I and $699.7\times$ even in the most demanding D-II + P-III case.  

\begin{table}[b]
  \caption{Average power and latency across configurations.
  The static powers are 150W (CPU for Pandas), 33W (RTX 3090), 43W (A100), and 17W (\textsc{PipeRec}). }
  \vspace{-0.5em}
  \label{tab:power}
  \centering
  \scalebox{0.85}{
  \begin{tblr}{
    colspec={ccccccccc},
    row{1} = {font=\bfseries},
    row{2} = {font=\bfseries},
    cell{1}{1} = {r=2}{c}, 
    cell{1}{2} = {c=2}{c}, 
    cell{1}{4} = {c=2}{c}, 
    cell{1}{6} = {c=2}{c}, 
    cell{1}{8} = {c=2}{c}, 
  }
  \hline
  \textbf{Config.} & \textbf{CPU} && \textbf{RTX 3090} && \textbf{A100} && \textbf{\textsc{PipeRec}} &\\
   & \textbf{Pwr.} & \textbf{Lat.} & \textbf{Pwr.} & \textbf{Lat.} & \textbf{Pwr.} & \textbf{Lat.} & \textbf{Pwr.} & \textbf{Lat.} \\
  \hline
  D-\URoman{1} + P-\URoman{1} & 294W & 78s & 92W  & 4.2s & 76W & 2.8s & 24W & 1.1s \\
  \emph{Eff. (CPU=1)} & \SetCell[c=2]{c}{1.0$\times$} && \SetCell[c=2]{c}{59.4$\times$} && \SetCell[c=2]{c}{107.8$\times$} && \SetCell[c=2]{c}{868.6$\times$} \\
  \hline
  D-\URoman{1} + P-\URoman{2} & 294W & 94s & 124W & 12.8s & 82W & 11.9s & 25W & 3.0s \\
  \emph{Eff. (CPU=1)} & \SetCell[c=2]{c}{1.0$\times$} && \SetCell[c=2]{c}{17.4$\times$} && \SetCell[c=2]{c}{28.3$\times$} && \SetCell[c=2]{c}{368.5$\times$} \\
  \hline
  D-\URoman{1} + P-\URoman{3} & 313W & 218s & 143W & 66.7s & 78W & 77.2s & 26W & 5.1s \\
  \emph{Eff. (CPU=1)} & \SetCell[c=2]{c}{1.0$\times$} && \SetCell[c=2]{c}{7.15$\times$} && \SetCell[c=2]{c}{11.3$\times$} && \SetCell[c=2]{c}{514.6$\times$} \\
  \hline
  D-\URoman{2} + P-\URoman{1} & 371W & 57s & 99W & 8.3s & 75W & 9.7s & 24W & 0.8s \\
  \emph{Eff. (CPU=1)} & \SetCell[c=2]{c}{1.0$\times$} && \SetCell[c=2]{c}{25.7$\times$} && \SetCell[c=2]{c}{29.1$\times$} && \SetCell[c=2]{c}{1101.4$\times$} \\
  \hline
  D-\URoman{2} + P-\URoman{2} & 363W & 61s & 113W & 15.4s & 75W & 16.7s & 25W & 1.5s \\
  \emph{Eff. (CPU=1)} & \SetCell[c=2]{c}{1.0$\times$} && \SetCell[c=2]{c}{12.7$\times$} && \SetCell[c=2]{c}{17.7$\times$} && \SetCell[c=2]{c}{590.5$\times$} \\
  \hline
  D-\URoman{2} + P-\URoman{3} & 379W & 72s & 119W & 25.8s & 76W & 24.4s & 26W & 1.5s \\
  \emph{Eff. (CPU=1)} & \SetCell[c=2]{c}{1.0$\times$} && \SetCell[c=2]{c}{8.9$\times$} && \SetCell[c=2]{c}{14.7$\times$} && \SetCell[c=2]{c}{699.7$\times$} \\
  \hline
  \end{tblr}}
\end{table}

These results highlight that while GPUs provide considerable improvements over CPUs, \textsc{PipeRec} achieves an order-of-magnitude higher Perf/W across both small and large pipelines.  
Thus, \textsc{PipeRec} accelerates preprocessing while also delivering superior energy efficiency, making it well suited for large-scale and power-constrained deployments.

\subsection{Resource Utilization}\label{sec:resource_utilization}
We further examine the hardware cost of \textsc{PipeRec} by analyzing the FPGA resource utilization.  
Table~\ref{tab:resource_fpga} reports the usage of configurable logic blocks (CLB), block RAM (BRAM), and DSP slices for three ETL pipelines (P-I, P-II, and P-III), the full-duplex RDMA stack, and their RDMA-enabled integration (R-P-I, R-P-II, and R-P-III).  
Here we neglect those resources unused.

The three pipelines exhibit moderate logic consumption, ranging from 17.6\% to 26.9\% CLB usage, and limited memory usage except for P-III, which requires 24.5\% BRAM due to the larger vocabulary table.  
DSP demand remains negligible, with only P-II and P-III consuming about 2.3\%.  
The standalone RDMA stack occupies 40.6\% of CLBs and 20.5\% of BRAM but does not require DSPs.  
When combined, the RDMA-enabled pipelines (R-P-I to R-P-III) inherit both the pipeline and RDMA resource demands.  
CLB utilization increases to 44.1--52.4\%, while BRAM usage remains modest (21.3--26.3\%).  
DSP usage stays constant at 2.3\%.  
Overall, even in the most demanding configuration (R-P-III), the design consumes just over half of the available CLBs and about one quarter of BRAM, leaving sufficient headroom for scaling or integration with additional components.  
These results demonstrate that \textsc{PipeRec} can be efficiently implemented on modern FPGAs: the pipelines themselves require lightweight resources, and the inclusion of a full-duplex RDMA stack incurs acceptable overheads while enabling end-to-end, high-throughput data delivery.

\begin{table}[t]
    \caption{Resource utilization for three pipelines, the full-duplex RDMA stack, and the corresponding RDMA-enabled pipelines (denoted as R-P-I, R-P-II, and R-P-III, where "R" indicates RDMA) in \textsc{PipeRec}.}
    \label{tab:resource_fpga}
    \centering
    \scalebox{0.85}{
    \begin{tblr}{
      colspec={ccccccc},  
    }
    \hline 
    Config  & P-\URoman{1} & P-\URoman{2} & P-\URoman{3} & RDMA & R-P-\URoman{1} & R-P-\URoman{2} & R-P-\URoman{3} \\
    \hline
    CLB & 17.6\% & 21.0\%  & 26.9\%  & 40.6\% & 44.1\%  & 45.5\%  & 52.4\% \\ 
    BRAM & 9.9\%  & 10.0\%  & 24.5\%  & 20.5\% & 21.3\% & 21.7\% & 26.3\%  \\ 
    DSP & 0.04\% & 2.3\% & 2.3\% & 0.0\% & 2.3\% & 2.3\%  & 2.3\%  \\
    \hline
    \end{tblr}
    }
\end{table}

\subsection{Concurrent Pipelines}\label{sec:scalability}
\textsc{PipeRec} supports the deployment of independent pipelines on the same device to enable higher parallelism or to serve different use cases simultaneously. 
We adjust the interface of dynamic regions provided by Coyote \cite{korolija2020abstractions}, which act as wrappers around each pipeline and enable spatial parallelism, thereby improving scalability. For instance, the Alveo V80 card \cite{alveo_v80} supports up to 800 Gbit/s of network throughput, which can be fully exploited by running multiple lightweight pipelines on a single board.

\begin{figure}[t]
    \centering 
    \includegraphics[width=0.6\linewidth]{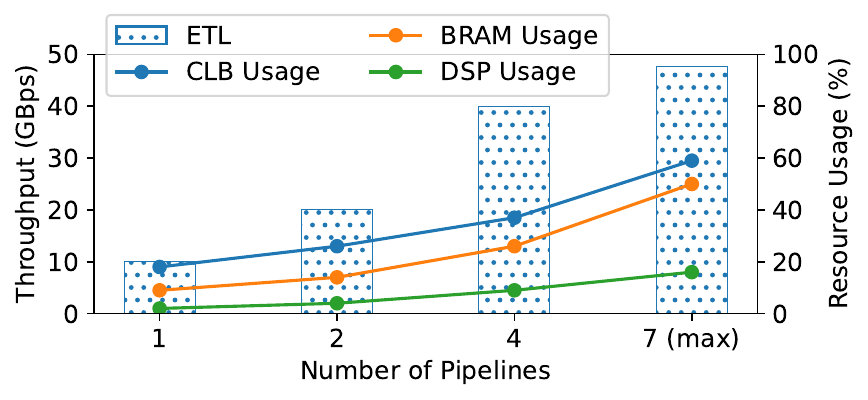}
    \caption{Throughput and resource utilization when multiple pipelines running concurrently. \textsc{PipeRec} now can serve up to 7 pipelines simultaneously.
    }
    \label{fig:multi_pipeline}
\end{figure}

To demonstrate \textsc{PipeRec}'s scalability, we evaluate its ability to linearly scale performance through the deployment of multiple pipelines. Specifically, we experiment with Pipeline I and Dataset II, which comprises 504 dense features and 42 sparse features. We deploy 1, 2, 4, and 7 pipelines, with 7 being the maximum number of dynamic regions fitting in the board we use for the prototype. Figure~\ref{fig:multi_pipeline} shows the resulting throughput, data loading speed, and resource utilization. The throughput scales linearly up to 4 pipelines, accompanied by a nearly linear increase in resource usage. Our prototype supports up to 7 concurrently running pipelines, albeit at a reduced clock frequency of 150MHz, which still matches the available network and PCIe bandwidth.

\section{Related Work}\label{sec:related}

\textbf{Preprocessing Service}.
Modern machine learning frameworks, such as \textit{tf.data} \cite{murray2021tf} and PyTorch's \textit{Dataloader} \cite{pytorch_dataloader}, optimize techniques like prefetching and streaming-like data loading. 
Researchers have focused on improving preprocessing performance at the software level, primarily on the CPU side \cite{phani2022uplift, kuchnik2022plumber, graur2022cachew, um2023fastflow, kim2023fusionflow, ye2025sand}.
In cloud environments, DPP \cite{zhao2022understanding} and GoldMiner \cite{zhao2023goldminer} highlight distributed solutions for preprocessing tasks within end-to-end training workflows.  
We use such multi-core CPU based systems as baselines for evaluating the performance of \textsc{PipeRec}.

\textbf{In-Network Acceleration}.
Network devices are extensively deployed in cloud environments \cite{nvidia_connectx, nvidia_bluefield, amd_pensando, alveo_u25, firestone2018azure}, playing a crucial role in facilitating efficient communication among nodes. 
Offloading CPU-intensive tasks, such as compression or decompression, to SmartNICs offers a promising solution to enhance the performance of distributed systems \cite{wei2023characterizing, kim2021linefs, tork2020lynx, ji2023styx}. 
Studies like FairNIC \cite{grant2020smartnic} and OSMOSIS \cite{khalilov2024osmosis} thoroughly analyze the multi-tenant capabilities in the network paths. 
We design and implement \textsc{PipeRec} to demonstrate the benefits of integrating preprocessing services within on-path SmartNICs to support the training process.

\textbf{P2P Communication}.
The paradigm of Peer-to-Peer (P2P) communication is popular in heterogeneous or distributed systems, where data can exchange directly between peers without passing through a centralized hub.
Prior research \cite{thoma2013fpga, bittner2014direct} has implemented GPUDirect RDMA on an FPGA to facilitate direct access to GPU memory, while FpgaNIC \cite{wang_atc22} further enhances this capability by enabling GPUs to trigger doorbell registers within an FPGA.


\section{Conclusion}\label{sec:conclusion}

ETL is the dominant bottleneck in modern recommender training rather than model compute. 
\textsc{PipeRec} addresses this with a hardware–accelerated, training-aware ETL engine: software-defined operators are compiled to reconfigurable FPGA pipelines; a co-scheduling runtime overlaps ETL with GPU training; and a vFPGA I/O stack streams training-ready batches directly to GPUs via P2P DMA, avoiding CPU staging.
Across diverse datasets, \textsc{PipeRec} accelerates ETL by up to two orders of magnitude over CPU systems and substantially over state-of-the-art GPU ETL, sustaining high GPU utilization and shortening end-to-end training time. It supports both stateless and stateful pipelines and scales through concurrent pipeline instantiation.

Beyond recommenders, the same hardware-accelerated ETL substrate generalizes to data-intensive online services (e.g., streaming analytics, telemetry, fraud/ad ranking, LLM inference service) where low-latency, high-throughput preprocessing governs cost and efficiency. \textsc{PipeRec} offers a concise blueprint for integrating training-aware ETL with accelerators at scale.






\bibliographystyle{ACM-Reference-Format}
\bibliography{sample}

@String{Computing = "Computing" }

@String{Computer = "{IEEE} Computer" }

@String{Springer = "Springer-Verlag" }

@inproceedings{ye2025sand,
  title={SAND: A New Programming Abstraction for Video-based Deep Learning},
  author={Ye, Juncheol and Lee, Seungkook and Lim, Hwijoon and Lee, Jihyuk and Hong, Uitaek and Kwon, Youngjin and Han, Dongsu},
  booktitle={Proceedings of the ACM SIGOPS 31st Symposium on Operating Systems Principles},
  pages={589--605},
  year={2025}
}

@inproceedings{thoma2013fpga,
  title={FPGA 2: An open source framework for FPGA-GPU PCIe communication},
  author={Thoma, Yann and Dassatti, Alberto and Molla, Daniel},
  booktitle={2013 international conference on reconfigurable computing and FPGAs (ReConFig)},
  pages={1--6},
  year={2013},
  organization={IEEE}
}

@article{bittner2014direct,
  title={Direct GPU/FPGA communication via PCI express},
  author={Bittner, Ray and Ruf, Erik and Forin, Alessandro},
  journal={Cluster Computing},
  volume={17},
  pages={339--348},
  year={2014},
  publisher={Springer}
}

@inproceedings{firestone2018azure,
  title={Azure accelerated networking: Smartnics in the public cloud},
  author={Firestone, Daniel and Putnam, Andrew and Mundkur, Sambhrama and Chiou, Derek and Dabagh, Alireza and Andrewartha, Mike and Angepat, Hari and Bhanu, Vivek and Caulfield, Adrian and Chung, Eric and others},
  booktitle={15th $\{$USENIX$\}$ Symposium on Networked Systems Design and Implementation ($\{$NSDI$\}$ 18)},
  pages={51--66},
  year={2018}
}

@misc{s-tui,
  author    = "The Stress Terminal UI: s-tui",
  title     = "\url{https://github.com/amanusk/s-tui}",
  year      = "2024"
}

@inproceedings{sima2022ekko,
  title={Ekko: A $\{$Large-Scale$\}$ deep learning recommender system with $\{$Low-Latency$\}$ model update},
  author={Sima, Chijun and Fu, Yao and Sit, Man-Kit and Guo, Liyi and Gong, Xuri and Lin, Feng and Wu, Junyu and Li, Yongsheng and Rong, Haidong and Aublin, Pierre-Louis and others},
  booktitle={16th USENIX Symposium on Operating Systems Design and Implementation (OSDI 22)},
  pages={821--839},
  year={2022}
}

@inproceedings{markidis2018nvidia,
  title={Nvidia tensor core programmability, performance \& precision},
  author={Markidis, Stefano and Der Chien, Steven Wei and Laure, Erwin and Peng, Ivy Bo and Vetter, Jeffrey S},
  booktitle={2018 IEEE international parallel and distributed processing symposium workshops (IPDPSW)},
  pages={522--531},
  year={2018},
  organization={IEEE}
}

@article{cherubin2020tools,
  title={Tools for reduced precision computation: a survey},
  author={Cherubin, Stefano and Agosta, Giovanni},
  journal={ACM Computing Surveys (CSUR)},
  volume={53},
  number={2},
  pages={1--35},
  year={2020},
  publisher={ACM New York, NY, USA}
}

@inproceedings{nandakumar2018mixed,
  title={Mixed-precision architecture based on computational memory for training deep neural networks},
  author={Nandakumar, SR and Le Gallo, Manuel and Boybat, Irem and Rajendran, Bipin and Sebastian, Abu and Eleftheriou, Evangelos},
  booktitle={2018 IEEE International Symposium on Circuits and Systems (ISCAS)},
  pages={1--5},
  year={2018},
  organization={IEEE}
}

@article{kalamkar2019study,
  title={A study of BFLOAT16 for deep learning training},
  author={Kalamkar, Dhiraj and Mudigere, Dheevatsa and Mellempudi, Naveen and Das, Dipankar and Banerjee, Kunal and Avancha, Sasikanth and Vooturi, Dharma Teja and Jammalamadaka, Nataraj and Huang, Jianyu and Yuen, Hector and others},
  journal={arXiv preprint arXiv:1905.12322},
  year={2019}
}

@article{sun2019hybrid,
  title={HFP8 training and inference for deep neural networks},
  author={Sun, Xiao and Choi, Jungwook and Chen, Chia-Yu and Wang, Naigang and Venkataramani, Swagath and Srinivasan, Vijayalakshmi Viji and Cui, Xiaodong and Zhang, Wei and Gopalakrishnan, Kailash},
  journal={Advances in neural information processing systems},
  volume={32},
  year={2019}
}

@article{boutros2024field,
  title={Field-Programmable Gate Array Architecture for Deep Learning: Survey \& Future Directions},
  author={Boutros, Andrew and Arora, Aman and Betz, Vaughn},
  journal={arXiv preprint arXiv:2404.10076},
  year={2024}
}

@inproceedings{khalilov2024osmosis,
  title={$\{$OSMOSIS$\}$: Enabling $\{$Multi-Tenancy$\}$ in Datacenter $\{$SmartNICs$\}$},
  author={Khalilov, Mikhail and Chrapek, Marcin and Shen, Siyuan and Vezzu, Alessandro and Benz, Thomas and Di Girolamo, Salvatore and Schneider, Timo and De Sensi, Daniele and Benini, Luca and Hoefler, Torsten},
  booktitle={2024 USENIX Annual Technical Conference (USENIX ATC 24)},
  pages={247--263},
  year={2024}
}

@misc{meta_dpp,
  author    = "Scaling data ingestion for machine learning training at Meta",
  title     = "\url{https://engineering.fb.com/2022/09/19/ml-applications/data-ingestion-machine-learning-training-meta/}",
  year      = "2022" 
}

@misc{xilinx_dma,
  author    = "Xilinx DMA/Bridge Subsystem for PCI Express v4.1",
  title     = "\url{https://www.amd.com/content/dam/xilinx/support/documents/ip_documentation/xdma/v4_1/pg195-pcie-dma.pdf}",
  year      = "2024" 
}

@misc{intel_dma,
  author    = "Intel Multichannel DMA Intel FPGA IP for PCI Express",
  title     = "\url{https://www.intel.com/content/www/us/en/products/details/fpga/intellectual-property/interface-protocols/multichannel-dma-mcdma.html}",
  year      = "2024" 
}

@misc{criteo_1T,
  author    = "Criteo 1TB Dataset",
  title     = "\url{https://ailab.criteo.com/download-criteo-1tb-click-logs-dataset}",
  year      = "2024" 
}

@misc{tf_dlrm,
  author    = "Tensorflow DLRM",
  title     = "\url{https://github.com/tensorflow/models/tree/master/official/recommendation/ranking}",
  year      = "2024" 
}

@misc{meta_dlrm,
  author    = "Meta's DLRM Preprocessing Pipeline",
  title     = "\url{https://github.com/facebookresearch/dlrm}",
  year      = "2024" 
}

@misc{criteo,
  author    = "Criteo Kaggle Dataset",
  title     = "\url{https://www.kaggle.com/datasets/mrkmakr/criteo-dataset}",
  year      = "2024" 
}

@misc{alveo_u55c,
  author    = "Xilinx Alveo U55c",
  title     = "\url{https://www.amd.com/en/products/accelerators/alveo/u55c/a-u55c-p00g-pq-g.html}",
  year      = "2024" 
}

@misc{gpu_direct_storage,
  author    = "Nvidia GPUDirect Storage",
  title     = "\url{https://developer.nvidia.com/blog/gpudirect-storage/}",
  year      = "2024" 
}

@article{hancock2020survey,
  title={Survey on categorical data for neural networks},
  author={Hancock, John T and Khoshgoftaar, Taghi M},
  journal={Journal of big data},
  volume={7},
  number={1},
  pages={28},
  year={2020},
  publisher={Springer}
}

@inproceedings{korolija2020abstractions,
  title={Do $\{$OS$\}$ abstractions make sense on $\{$FPGAs$\}$?},
  author={Korolija, Dario and Roscoe, Timothy and Alonso, Gustavo},
  booktitle={14th USENIX Symposium on Operating Systems Design and Implementation (OSDI 20)},
  pages={991--1010},
  year={2020}
}

@article{jiang2021microrec,
  title={MicroRec: efficient recommendation inference by hardware and data structure solutions},
  author={Jiang, Wenqi and He, Zhenhao and Zhang, Shuai and Preu{\ss}er, Thomas B and Zeng, Kai and Feng, Liang and Zhang, Jiansong and Liu, Tongxuan and Li, Yong and Zhou, Jingren and others},
  journal={Proceedings of Machine Learning and Systems},
  volume={3},
  pages={845--859},
  year={2021}
}

@inproceedings{jiang2021fleetrec,
  title={Fleetrec: Large-scale recommendation inference on hybrid gpu-fpga clusters},
  author={Jiang, Wenqi and He, Zhenhao and Zhang, Shuai and Zeng, Kai and Feng, Liang and Zhang, Jiansong and Liu, Tongxuan and Li, Yong and Zhou, Jingren and Zhang, Ce and others},
  booktitle={Proceedings of the 27th ACM SIGKDD Conference on Knowledge Discovery \& Data Mining},
  pages={3097--3105},
  year={2021}
}

@inproceedings{zhu2021distributed,
  title={Distributed recommendation inference on fpga clusters},
  author={Zhu, Yu and He, Zhenhao and Jiang, Wenqi and Zeng, Kai and Zhou, Jingren and Alonso, Gustavo},
  booktitle={2021 31st International Conference on Field-Programmable Logic and Applications (FPL)},
  pages={279--285},
  year={2021},
  organization={IEEE}
}

@article{zhao2023goldminer,
  title={GoldMiner: Elastic Scaling of Training Data Pre-Processing Pipelines for Deep Learning},
  author={Zhao, Hanyu and Yang, Zhi and Cheng, Yu and Tian, Chao and Ren, Shiru and Xiao, Wencong and Yuan, Man and Chen, Langshi and Liu, Kaibo and Zhang, Yang and others},
  journal={Proceedings of the ACM on Management of Data},
  volume={1},
  number={2},
  pages={1--25},
  year={2023},
  publisher={ACM New York, NY, USA}
}

@inproceedings{zhao2022understanding,
  title={Understanding data storage and ingestion for large-scale deep recommendation model training: Industrial product},
  author={Zhao, Mark and Agarwal, Niket and Basant, Aarti and Gedik, Bu{\u{g}}ra and Pan, Satadru and Ozdal, Mustafa and Komuravelli, Rakesh and Pan, Jerry and Bao, Tianshu and Lu, Haowei and others},
  booktitle={Proceedings of the 49th annual international symposium on computer architecture},
  pages={1042--1057},
  year={2022}
}

@article{qi2023auto,
  title={Auto-FP: An Experimental Study of Automated Feature Preprocessing for Tabular Data},
  author={Qi, Danrui and Peng, Jinglin and He, Yongjun and Wang, Jiannan},
  journal={arXiv preprint arXiv:2310.02540},
  year={2023}
}

@inproceedings{li2021cleanml,
  title={Cleanml: A study for evaluating the impact of data cleaning on ml classification tasks},
  author={Li, Peng and Rao, Xi and Blase, Jennifer and Zhang, Yue and Chu, Xu and Zhang, Ce},
  booktitle={2021 IEEE 37th International Conference on Data Engineering (ICDE)},
  pages={13--24},
  year={2021},
  organization={IEEE}
}

@article{li2023volcanoml,
  title={VolcanoML: speeding up end-to-end AutoML via scalable search space decomposition},
  author={Li, Yang and Shen, Yu and Zhang, Wentao and Zhang, Ce and Cui, Bin},
  journal={The VLDB Journal},
  volume={32},
  number={2},
  pages={389--413},
  year={2023},
  publisher={Springer}
}

@article{he2021automl,
  title={AutoML: A survey of the state-of-the-art},
  author={He, Xin and Zhao, Kaiyong and Chu, Xiaowen},
  journal={Knowledge-based systems},
  volume={212},
  pages={106622},
  year={2021},
  publisher={Elsevier}
}

@article{budach2022effects,
  title={The effects of data quality on machine learning performance},
  author={Budach, Lukas and Feuerpfeil, Moritz and Ihde, Nina and Nathansen, Andrea and Noack, Nele and Patzlaff, Hendrik and Naumann, Felix and Harmouch, Hazar},
  journal={arXiv preprint arXiv:2207.14529},
  year={2022}
}

@article{chen2021data,
  title={Data evaluation and enhancement for quality improvement of machine learning},
  author={Chen, Haihua and Chen, Jiangping and Ding, Junhua},
  journal={IEEE Transactions on Reliability},
  volume={70},
  number={2},
  pages={831--847},
  year={2021},
  publisher={IEEE}
}

@inproceedings{jain2020overview,
  title={Overview and importance of data quality for machine learning tasks},
  author={Jain, Abhinav and Patel, Hima and Nagalapatti, Lokesh and Gupta, Nitin and Mehta, Sameep and Guttula, Shanmukha and Mujumdar, Shashank and Afzal, Shazia and Sharma Mittal, Ruhi and Munigala, Vitobha},
  booktitle={Proceedings of the 26th ACM SIGKDD international conference on knowledge discovery \& data mining},
  pages={3561--3562},
  year={2020}
}

@article{naumov2019deep,
  title={Deep learning recommendation model for personalization and recommendation systems},
  author={Naumov, Maxim and Mudigere, Dheevatsa and Shi, Hao-Jun Michael and Huang, Jianyu and Sundaraman, Narayanan and Park, Jongsoo and Wang, Xiaodong and Gupta, Udit and Wu, Carole-Jean and Azzolini, Alisson G and others},
  journal={arXiv preprint arXiv:1906.00091},
  year={2019}
}

@inproceedings{oldridge2020merlin,
  title={Merlin: a gpu accelerated recommendation framework},
  author={Oldridge, Even and Perez, Julio and Frederickson, Ben and Koumchatzky, Nicolas and Lee, Minseok and Wang, Zehuan and Wu, Lei and Yu, Fan and Zamora, Rick and Yilmaz, Onur and others},
  booktitle={Proceedings of IRS},
  year={2020}
}

@inproceedings{gupta2020architectural,
  title={The architectural implications of facebook's dnn-based personalized recommendation},
  author={Gupta, Udit and Wu, Carole-Jean and Wang, Xiaodong and Naumov, Maxim and Reagen, Brandon and Brooks, David and Cottel, Bradford and Hazelwood, Kim and Hempstead, Mark and Jia, Bill and others},
  booktitle={2020 IEEE International Symposium on High Performance Computer Architecture (HPCA)},
  pages={488--501},
  year={2020},
  organization={IEEE}
}

@article{phani2022uplift,
  title={UPLIFT: parallelization strategies for feature transformations in machine learning workloads},
  author={Phani, Arnab and Erlbacher, Lukas and Boehm, Matthias},
  journal={Proceedings of the VLDB Endowment},
  volume={15},
  number={11},
  pages={2929--2938},
  year={2022},
  publisher={VLDB Endowment}
}

@inproceedings{sidler2020strom,
  title={StRoM: smart remote memory},
  author={Sidler, David and Wang, Zeke and Chiosa, Monica and Kulkarni, Amit and Alonso, Gustavo},
  booktitle={Proceedings of the Fifteenth European Conference on Computer Systems},
  pages={1--16},
  year={2020}
}

@article{kuchnik2022plumber,
  title={Plumber: Diagnosing and removing performance bottlenecks in machine learning data pipelines},
  author={Kuchnik, Michael and Klimovic, Ana and Simsa, Jiri and Smith, Virginia and Amvrosiadis, George},
  journal={Proceedings of Machine Learning and Systems},
  volume={4},
  pages={33--51},
  year={2022}
}

@article{murray2021tf,
  title={tf.data: A machine learning data processing framework},
  author={Murray, Derek G and Simsa, Jiri and Klimovic, Ana and Indyk, Ihor},
  journal={arXiv preprint arXiv:2101.12127},
  year={2021}
}

@article{audibert2023tf,
  title={tf.data service: A Case for Disaggregating ML Input Data Processing},
  author={Audibert, Andrew and Chen, Yang and Graur, Dan and Klimovic, Ana and {\v{S}}im{\v{s}}a, Ji{\v{r}}{\'\i} and Chandramohan, A},
  year={2023}
}

@inproceedings{graur2022cachew,
  title={Cachew: Machine learning input data processing as a service},
  author={Graur, Dan and Aymon, Damien and Kluser, Dan and Albrici, Tanguy and Thekkath, Chandramohan A and Klimovic, Ana},
  booktitle={2022 USENIX Annual Technical Conference (USENIX ATC 22)},
  pages={689--706},
  year={2022}
}

@inproceedings{mudigere2022software,
  title={Software-hardware co-design for fast and scalable training of deep learning recommendation models},
  author={Mudigere, Dheevatsa and Hao, Yuchen and Huang, Jianyu and Jia, Zhihao and Tulloch, Andrew and Sridharan, Srinivas and Liu, Xing and Ozdal, Mustafa and Nie, Jade and Park, Jongsoo and others},
  booktitle={Proceedings of the 49th Annual International Symposium on Computer Architecture},
  pages={993--1011},
  year={2022}
}

@article{choquette2021nvidia,
  title={Nvidia a100 tensor core gpu: Performance and innovation},
  author={Choquette, Jack and Gandhi, Wishwesh and Giroux, Olivier and Stam, Nick and Krashinsky, Ronny},
  journal={IEEE Micro},
  volume={41},
  number={2},
  pages={29--35},
  year={2021},
  publisher={IEEE}
}

@article{choquette2023nvidia,
  title={Nvidia hopper h100 gpu: Scaling performance},
  author={Choquette, Jack},
  journal={IEEE Micro},
  volume={43},
  number={3},
  pages={9--17},
  year={2023},
  publisher={IEEE}
}

@inproceedings{tirumala2024nvidia,
  title={NVIDIA Blackwell Platform: Advancing Generative AI and Accelerated Computing},
  author={Tirumala, Ajay and Wong, Raymond},
  booktitle={2024 IEEE Hot Chips 36 Symposium (HCS)},
  pages={1--33},
  year={2024},
  organization={IEEE Computer Society}
}

@inproceedings{fan2004gpu,
  title={GPU cluster for high performance computing},
  author={Fan, Zhe and Qiu, Feng and Kaufman, Arie and Yoakum-Stover, Suzanne},
  booktitle={SC'04: Proceedings of the 2004 ACM/IEEE conference on Supercomputing},
  pages={47--47},
  year={2004},
  organization={IEEE}
}

@misc{nvtabular_blog,
  author    = "NVIDIA NVTabular Technical Blog",
  title     = "\url{https://developer.nvidia.com/blog/announcing-the-nvtabular-open-beta-with-multi-gpu-support-and-new-data-loaders/}",
  year      = "2020" 
}

@misc{hadoop,
  author    = "Apache Hadoop Distributed File System",
  title     = "\url{https://hadoop.apache.org/}",
  year      = "2024" 
}

@misc{amazon_s3,
  author    = "Amazon S3 for Cloud Storage",
  title     = "\url{https://aws.amazon.com/s3/}",
  year      = "2024" 
}

@misc{nvidia_bluefield,
  author    = "Nvidia Bluefield Networking Platform",
  title     = "\url{https://www.nvidia.com/en-us/networking/products/data-processing-unit/}",
  year      = "2024" 
}

@misc{nvidia_connectx,
  author    = "Nvidia ConnextX NICs",
  title     = "\url{https://www.nvidia.com/en-us/networking/ethernet-adapters/}",
  year      = "2024"
}

@misc{amd_pensando,
  author    = "AMD Pensando Networking",
  title     = "\url{https://www.amd.com/en/products/accelerators/pensando.html}",
  year      = "2024"
}

@misc{alveo_u25,
  author    = "Xilinx Alveo U25",
  title     = "\url{https://www.xilinx.com/publications/product-briefs/alveo-u25-product-brief.pdf}",
  year      = "2024"
}

@article{zhao2021understanding,
  title={Understanding and co-designing the data ingestion pipeline for industry-scale recsys training},
  author={Zhao, Mark and Agarwal, Niket and Basant, Aarti and Gedik, Bugra and Pan, Satadru and Ozdal, Mustafa and Komuravelli, Rakesh and Pan, Jerry and Bao, Tianshu and Lu, Haowei and others},
  journal={arXiv preprint arXiv:2108.09373},
  volume={1},
  number={3.1},
  pages={2},
  year={2021}
}

@article{zhang2020understanding,
  title={Understanding the effect of data center resource disaggregation on production dbmss},
  author={Zhang, Qizhen and Cai, Yifan and Chen, Xinyi and Angel, Sebastian and Chen, Ang and Liu, Vincent and Loo, Boon Thau},
  journal={Proceedings of the VLDB Endowment},
  volume={13},
  number={9},
  year={2020}
}

@article{liu2022monolith,
  title={Monolith: real time recommendation system with collisionless embedding table},
  author={Liu, Zhuoran and Zou, Leqi and Zou, Xuan and Wang, Caihua and Zhang, Biao and Tang, Da and Zhu, Bolin and Zhu, Yijie and Wu, Peng and Wang, Ke and others},
  journal={RecSys},
  year={2022}
}

@article{um2023fastflow,
  title={Fastflow: Accelerating deep learning model training with smart offloading of input data pipeline},
  author={Um, Taegeon and Oh, Byungsoo and Seo, Byeongchan and Kweun, Minhyeok and Kim, Goeun and Lee, Woo-Yeon},
  journal={Proceedings of the VLDB Endowment},
  volume={16},
  number={5},
  pages={1086--1099},
  year={2023},
  publisher={VLDB Endowment}
}

@inproceedings{wang_atc22,
  title={FpgaNIC: An FPGA-based Versatile 100Gb SmartNIC for GPUs},
  author={Zeke Wang and Hongjing Huang and Jie Zhang and Fei Wu and Gustavo Alonso},
  year={2022},
  booktitle={2022 USENIX Annual Technical Conference (ATC)},
}

@inproceedings{he2024accl+,
  title={$\{$ACCL+$\}$: an $\{$FPGA-Based$\}$ Collective Engine for Distributed Applications},
  author={He, Zhenhao and Korolija, Dario and Zhu, Yu and Ramhorst, Benjamin and Laan, Tristan and Petrica, Lucian and Blott, Michaela and Alonso, Gustavo},
  booktitle={18th USENIX Symposium on Operating Systems Design and Implementation (OSDI 24)},
  pages={211--231},
  year={2024}
}

@inproceedings{qin2024datasetgrowth,
      title={Dataset Growth}, 
      author={Ziheng Qin and Zhaopan Xu and Yukun Zhou and Zangwei Zheng and Zebang Cheng and Hao Tang and Lei Shang and Baigui Sun and Xiaojiang Peng and Radu Timofte and Hongxun Yao and Kai Wang and Yang You},
      booktitle={ECCV},
      year={2024}
}

@inproceedings{kim2021linefs,
  title={LineFS: Efficient SmartNIC offload of a distributed file system with pipeline parallelism},
  author={Kim, Jongyul and Jang, Insu and Reda, Waleed and Im, Jaeseong and Canini, Marco and Kosti{\'c}, Dejan and Kwon, Youngjin and Peter, Simon and Witchel, Emmett},
  booktitle={Proceedings of the ACM SIGOPS 28th Symposium on Operating Systems Principles},
  pages={756--771},
  year={2021}
}

@misc{nvidia_nvtabular_dlrm,
  author    = "Accelerating ETL for Recommender Systems on NVIDIA GPUs with NVTabular",
  title     = "\url{https://developer.nvidia.com/blog/accelerating-etl-for-recsys-on-gpus-with-nvtabular/}",
  year      = "2020" 
}

@misc{eirinaki2018recommender,
  title={Recommender systems for large-scale social networks: A review of challenges and solutions},
  author={Eirinaki, Magdalini and Gao, Jerry and Varlamis, Iraklis and Tserpes, Konstantinos},
  journal={Future generation computer systems},
  volume={78},
  pages={413--418},
  year={2018},
  publisher={Elsevier}
}

@article{song2014online,
  title={Online learning in large-scale contextual recommender systems},
  author={Song, Linqi and Tekin, Cem and Van Der Schaar, Mihaela},
  journal={IEEE Transactions on Services Computing},
  volume={9},
  number={3},
  pages={433--445},
  year={2014},
  publisher={IEEE}
}

@inproceedings{freno2017practical,
  title={Practical lessons from developing a large-scale recommender system at Zalando},
  author={Freno, Antonino},
  booktitle={Proceedings of the eleventh ACM conference on recommender systems},
  pages={251--259},
  year={2017}
}

@inproceedings{zhou2021contrastive,
  title={Contrastive learning for debiased candidate generation in large-scale recommender systems},
  author={Zhou, Chang and Ma, Jianxin and Zhang, Jianwei and Zhou, Jingren and Yang, Hongxia},
  booktitle={Proceedings of the 27th ACM SIGKDD Conference on Knowledge Discovery \& Data Mining},
  pages={3985--3995},
  year={2021}
}

@article{guo2023evaluating,
  title={Evaluating online bandit exploration in large-scale recommender system},
  author={Guo, Hongbo and Naeff, Ruben and Nikulkov, Alex and Zhu, Zheqing},
  journal={arXiv preprint arXiv:2304.02572},
  year={2023}
}

@inproceedings{tork2020lynx,
  title={Lynx: A smartnic-driven accelerator-centric architecture for network servers},
  author={Tork, Maroun and Maudlej, Lina and Silberstein, Mark},
  booktitle={Proceedings of the Twenty-Fifth International Conference on Architectural Support for Programming Languages and Operating Systems},
  pages={117--131},
  year={2020}
}

@inproceedings{grant2020smartnic,
  title={Smartnic performance isolation with fairnic: Programmable networking for the cloud},
  author={Grant, Stewart and Yelam, Anil and Bland, Maxwell and Snoeren, Alex C},
  booktitle={Proceedings of the Annual conference of the ACM Special Interest Group on Data Communication on the applications, technologies, architectures, and protocols for computer communication},
  pages={681--693},
  year={2020}
}

@inproceedings{ji2023styx,
  title={$\{$STYX$\}$: Exploiting $\{$SmartNIC$\}$ Capability to Reduce Datacenter Memory Tax},
  author={Ji, Houxiang and Mansi, Mark and Sun, Yan and Yuan, Yifan and Huang, Jinghan and Kuper, Reese and Swift, Michael M and Kim, Nam Sung},
  booktitle={2023 USENIX Annual Technical Conference (USENIX ATC 23)},
  pages={619--633},
  year={2023}
}

@inproceedings{wei2023characterizing,
  title={Characterizing Off-path $\{$SmartNIC$\}$ for Accelerating Distributed Systems},
  author={Wei, Xingda and Cheng, Rongxin and Yang, Yuhan and Chen, Rong and Chen, Haibo},
  booktitle={17th USENIX Symposium on Operating Systems Design and Implementation (OSDI 23)},
  pages={987--1004},
  year={2023}
}

@article{ofeidis2022overview,
  title={An overview of the data-loader landscape: Comparative performance analysis},
  author={Ofeidis, Iason and Kiedanski, Diego and Tassiulas, Leandros},
  journal={arXiv preprint arXiv:2209.13705},
  year={2022}
}

@article{cai2020fly,
  title={On-the-fly data loader and utterance-level aggregation for speaker and language recognition},
  author={Cai, Weicheng and Chen, Jinkun and Zhang, Jun and Li, Ming},
  journal={IEEE/ACM Transactions on Audio, Speech, and Language Processing},
  volume={28},
  pages={1038--1051},
  year={2020},
  publisher={IEEE}
}

@article{bai2021efficient,
  title={Efficient data loader for fast sampling-based GNN training on large graphs},
  author={Bai, Youhui and Li, Cheng and Lin, Zhiqi and Wu, Yufei and Miao, Youshan and Liu, Yunxin and Xu, Yinlong},
  journal={IEEE Transactions on Parallel and Distributed Systems},
  volume={32},
  number={10},
  pages={2541--2556},
  year={2021},
  publisher={IEEE}
}

@inproceedings{jia2022data,
  title={A Data-Loader Tunable Knob to Shorten GPU Idleness for Distributed Deep Learning},
  author={Jia, Danlin and Yuan, Geng and Lin, Xue and Mi, Ningfang},
  booktitle={2022 IEEE 15th International Conference on Cloud Computing (CLOUD)},
  pages={449--458},
  year={2022},
  organization={IEEE}
}

@article{zolnouri2020importance,
  title={Importance of data loading pipeline in training deep neural networks},
  author={Zolnouri, Mahdi and Li, Xinlin and Nia, Vahid Partovi},
  journal={arXiv preprint arXiv:2005.02130},
  year={2020}
}

@article{svogor2022profiling,
  title={Profiling and improving the pytorch dataloader for high-latency storage: A technical report},
  author={Svogor, Ivan and Eichenberger, Christian and Spanring, Markus and Neun, Moritz and Kopp, Michael},
  journal={arXiv preprint arXiv:2211.04908},
  year={2022}
}

@misc{vitis_hls,
  author    = "Xilinx Vitis HLS",
  title     = "\url{https://www.amd.com/en/products/software/adaptive-socs-and-fpgas/vitis/vitis-hls.html}",
  year      = "2024" 
}

@misc{xilinx_xrt,
  author    = "Xilinx Runtime",
  title     = "\url{https://xilinx.github.io/XRT/master/html/index.html}",
  year      = "2024" 
}

@article{zhao2023embedding,
  title={Embedding in recommender systems: A survey},
  author={Zhao, Xiangyu and Wang, Maolin and Zhao, Xinjian and Li, Jiansheng and Zhou, Shucheng and Yin, Dawei and Li, Qing and Tang, Jiliang and Guo, Ruocheng},
  journal={arXiv preprint arXiv:2310.18608},
  year={2023}
}

@misc{alveo_v80,
  author    = "Xilinx Alveo V80",
  title     = "\url{https://www.amd.com/en/products/accelerators/alveo/v80.html}",
  year      = "2024" 
}

@article{dally2021evolution,
  title={Evolution of the graphics processing unit (GPU)},
  author={Dally, William J and Keckler, Stephen W and Kirk, David B},
  journal={IEEE Micro},
  volume={41},
  number={6},
  pages={42--51},
  year={2021},
  publisher={IEEE}
}

@article{kim2023fusionflow,
  title={FusionFlow: Accelerating Data Preprocessing for Machine Learning with CPU-GPU Cooperation},
  author={Kim, Taeyoon and Park, ChanHo and Mukimbekov, Mansur and Hong, Heelim and Kim, Minseok and Jin, Ze and Kim, Changdae and Shin, Ji-Yong and Jeon, Myeongjae},
  journal={Proceedings of the VLDB Endowment},
  volume={17},
  number={4},
  pages={863--876},
  year={2023},
  publisher={VLDB Endowment}
}

@inproceedings{cheng2016wide,
  title={Wide \& deep learning for recommender systems},
  author={Cheng, Heng-Tze and Koc, Levent and Harmsen, Jeremiah and Shaked, Tal and Chandra, Tushar and Aradhye, Hrishi and Anderson, Glen and Corrado, Greg and Chai, Wei and Ispir, Mustafa and others},
  booktitle={Proceedings of the 1st workshop on deep learning for recommender systems},
  pages={7--10},
  year={2016}
}

@misc{pytorch_dataloader,
  author    = "Pytorch Dataloader",
  title     = "\url{https://pytorch.org/tutorials/beginner/basics/data_tutorial.html}",
  year      = "2024" 
}

@misc{nvidia_dali,
  author    = "Nvidia Data Loading Library",
  title     = "\url{https://github.com/NVIDIA/DALI/tree/main/dali/pipeline}",
  year      = "2024" 
}

@misc{nvidia_rapids,
  author    = "Nvidia Rapids",
  title     = "\url{https://developer.nvidia.com/rapids}",
  year      = "2024" 
}

@article{vohra2016apache,
  title={Apache parquet},
  author={Vohra, Deepak and Vohra, Deepak},
  journal={Practical Hadoop Ecosystem: A Definitive Guide to Hadoop-Related Frameworks and Tools},
  pages={325--335},
  year={2016},
  publisher={Springer}
}

@inproceedings{lee2024presto,
  title={PreSto: An In-Storage Data Preprocessing System for Training Recommendation Models},
  author={Lee, Yunjae and Kim, Hyeseong and Rhu, Minsoo},
  booktitle={2024 ACM/IEEE 51st Annual International Symposium on Computer Architecture (ISCA)},
  pages={340--353},
  year={2024},
  organization={IEEE}
}

@inproceedings{Bother2025Modyn,
  author = {B\"{o}ther, Maximilian and Robroek, Ties and Gsteiger, Viktor and Ma, Xianzhe and T\"{o}z\"{u}n, P{\i}nar and Klimovic, Ana},
  title = {Modyn: Data-Centric Machine Learning Pipeline Orchestration},
  booktitle = {Proceedings of the Conference on Management of Data (SIGMOD)},
  year = {2025},
}

@inproceedings{egg2021online,
  title={Online learning for recommendations at grubhub},
  author={Egg, Alex},
  booktitle={Proceedings of the 15th ACM Conference on Recommender Systems},
  pages={569--571},
  year={2021}
}

@article{ramirez2017survey,
  title={A survey on data preprocessing for data stream mining: Current status and future directions},
  author={Ram{\'\i}rez-Gallego, Sergio and Krawczyk, Bartosz and Garc{\'\i}a, Salvador and Wo{\'z}niak, Micha{\l} and Herrera, Francisco},
  journal={Neurocomputing},
  volume={239},
  pages={39--57},
  year={2017},
  publisher={Elsevier}
}

@inproceedings{graur2024pecan,
  title={Pecan:$\{$Cost-Efficient$\}$$\{$ML$\}$ Data Preprocessing with Automatic Transformation Ordering and Hybrid Placement},
  author={Graur, Dan and Mraz, Oto and Li, Muyu and Pourghannad, Sepehr and Thekkath, Chandramohan A and Klimovic, Ana},
  booktitle={2024 USENIX Annual Technical Conference (USENIX ATC 24)},
  pages={649--665},
  year={2024}
}

@inproceedings{wang2024rap,
  title={Rap: Resource-aware automated gpu sharing for multi-gpu recommendation model training and input preprocessing},
  author={Wang, Zheng and Wang, Yuke and Deng, Jiaqi and Zheng, Da and Li, Ang and Ding, Yufei},
  booktitle={Proceedings of the 29th ACM International Conference on Architectural Support for Programming Languages and Operating Systems, Volume 2},
  pages={964--979},
  year={2024}
}










\end{document}